\pdfoutput=1   
\documentclass[cernpreprint,atlasdraft=true,texlive=2016,UKenglish,texmf]{atlasdoc}
\PreprintIdNumber{CERN-EP-2021-004}
 
\usepackage{atlaspackage}
\usepackage{atlasbiblatex}
 
\usepackage{atlasphysics}
 
\graphicspath{{logos/}{figures/}}
 
\usepackage{ANA-HDBS-2018-51-PAPER-defs}

\AtlasTitle{Search for charged Higgs bosons decaying into a top quark and a bottom quark at $\sqrt{s}$=13 TeV with the ATLAS detector}
 
\AtlasVersion{2.10}
\AtlasAbstract{
A search for charged Higgs bosons
decaying into a top quark and a bottom quark is presented.
The data analysed correspond to 139 fb$^{-1}$ of proton--proton collisions at $\sqrt{s}$=13 TeV,
recorded with the ATLAS detector at the LHC.
The production of a heavy charged Higgs boson in association with a top quark and a bottom quark,
$pp \rightarrow tbH^+ \rightarrow tbtb$,
is explored in the $H^+$ mass range from 200 to 2000~GeV using
final states with jets and one electron or muon.
Events are categorised according to the multiplicity of jets
and $b$-tagged jets,
and multivariate analysis techniques are used to discriminate between signal and background events.
No significant excess above the background-only hypothesis is observed
and exclusion limits are derived for the production cross-section times branching ratio of a
charged Higgs boson as a function of its mass;
they range from 3.6~pb at 200~GeV to 0.036~pb at 2000~GeV at 95\% confidence level.
The results are interpreted in the hMSSM and $M_h^{125}$ scenarios.
}
 
\author{The ATLAS Collaboration}
 
\AtlasRefCode{HDBS-2018-51}

\AtlasJournalRef{JHEP 06 (2021) 145}  
 
\PreprintIdNumber{CERN-EP-2021-004} 
 
\AtlasDOI{10.1007/JHEP06(2021)145}  
 
\hypersetup{pdftitle={ATLAS document},pdfauthor={The ATLAS Collaboration}}
 
\begin{document}
 
\maketitle
 
\tableofcontents   
 
\section{Introduction}
\begin{figure}[htb]
\centering
\includegraphics[width = 0.3\textwidth]{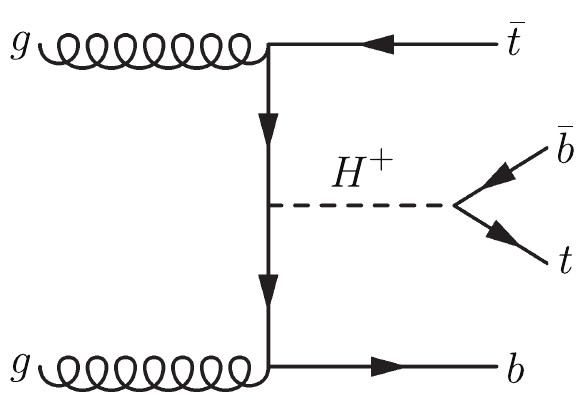}
\caption{
Leading-order Feynman diagram for the production of a heavy charged Higgs boson
in association with a top antiquark and a bottom quark, as well as its decay into a top quark and a bottom antiquark.}
\label{fig:Feynman}
\end{figure}

The discovery of a Higgs boson with a measured mass of 125~\GeV\
at the Large Hadron Collider (LHC) in 2012~\cite{HIGG-2012-27,CMS-HIG-12-028,Aad:2015zhl}
raises the question of whether this is the Higgs boson of the Standard Model (SM)
or part of an extended scalar sector.
Charged Higgs bosons\footnote{In the following,
charged Higgs bosons are denoted \hplus,
with the charge-conjugate $H^-$ always implied.
Similarly, the difference between quarks and antiquarks $q$ and $\bar q$ is generally understood from the context,
so that e.g.\ \hplustb\ means both $H^+ \rightarrow t \bar b$ and $H^- \rightarrow \bar t b$.}
are predicted in several extensions of the SM that add
a second doublet~\cite{Lee:1973iz,Branco:2011iw,Inoue:1982pi,Gunion:2002zf}
or triplets~\cite{trip1,trip2,trip3,trip4,trip5}
to the scalar sector.
In CP-conserving two-Higgs-doublet models (2HDMs),
the properties of the charged Higgs boson depend on its mass,
the mixing angle $\alpha$ of the neutral CP-even Higgs bosons,
and the ratio of the vacuum expectation values of the two Higgs doublets ($\tan \beta$).
This analysis searches for charged Higgs bosons
heavier than the top quark and
decaying into a top quark and a bottom quark.
At the LHC, charged Higgs bosons in this mass range are expected to be produced primarily in association with a top quark and a bottom quark~\cite{deFlorian:2016spz},
as illustrated in Figure~\ref{fig:Feynman}.
 
The ATLAS and CMS collaborations have searched for charged Higgs bosons
in proton--proton ($pp$) collisions at $\sqrt{s} = 7,8$ and 13~\TeV\
with data samples ranging from 2.9 to 36~\ifb,
probing the mass range below the top-quark mass in the
$\tau\nu$~\cite{HIGG-2012-09,HIGG-2012-22,HIGG-2013-30,CMS-HIG-11-019,CMS-HIG-14-023,HIGG-2016-11},
$cs$~\cite{HIGG-2012-10,CMS-HIG-13-035},
and $cb$~\cite{CMS-HIG-16-030} decay modes,
as well as above the top-quark mass in the $\tau\nu$ and $tb$ decay modes~\cite{HIGG-2013-30,CMS-HIG-14-023,HIGG-2013-28,HIGG-2015-11,HIGG-2016-11,HIGG-2017-04,CMS-HIG-18-004,CMS-HIG-18-015}.
In addition,
$H^+ \rightarrow WZ$ decays were searched for in the vector-boson-fusion production mode~\cite{HIGG-2014-13,CMS-HIG-16-027}.
ATLAS has also set limits on \hplus\ production in
a search for dijet resonances in events with an isolated lepton
using the Run~2 dataset~\cite{EXOT-2018-32}.
No evidence of charged Higgs bosons was found in any of these searches.
 
This paper presents an updated search for \hplus\ production in the \hplustb\ decay mode
with the full Run~2 dataset of $pp$ collisions taken at \comEne.
This decay mode has the highest branching ratio for charged Higgs bosons above the top-quark mass~\cite{Degrande:2015vpa}.
Events with one charged lepton ($\ell = e,\mu$) and jets in the final state are considered,
and exclusive regions are defined according to the overall number of jets,
and the number of jets tagged as containing a $b$-hadron.
In order to separate signal from SM background,
multivariate analysis (MVA) techniques combining jet multiplicities and several kinematic variables are employed
in the regions where the signal rate is expected to be largest.
Compared to the ATLAS result using 36~\ifb of Run 2 data~\cite{HIGG-2017-04},
improved limits on the $pp \rightarrow \tbhplus$ production cross-section times the \hplustb\ branching ratio
are set by means of a simultaneous fit
to the MVA classifier outputs in the different analysis regions,
which determines both the contribution from the \hplustb\ signal and the normalisation of the backgrounds.
The improvement is small at low \hplus\ mass,
where the measurement is dominated by systematic uncertainties,
but larger than the simple scaling with the square root of the ratio of integrated luminosities at high \hplus\ mass.
The results are interpreted in the framework of the
hMSSM~\cite{Djouadi:2013vqa, Maiani:2013hud, Djouadi:2013uqa, Djouadi:2015jea}
and various $M_h^{125}$ benchmark scenarios of the Minimal Supersymmetric Standard Model (MSSM)~\cite{Bagnaschi_2019,Degrande:2015vpa,Flechl_2015,deFlorian:2016spz,Dittmaier_2011,Berger_2005}.
 
\section{Data and simulation samples}
\label{sec:dataMC}
The data used in this analysis were recorded with the ATLAS detector
at the LHC between 2015 and 2018 from \comEne\ $pp$ collisions,
and correspond to an integrated luminosity of 139 \ifb.
ATLAS~\cite{PERF-2007-01, ATLAS-TDR-2010-19, PIX-2018-001} is a multipurpose detector
with a forward--backward symmetric cylindrical geometry and a near \(4\pi\) coverage in solid angle.\footnote{\AtlasCoordFootnote}
It consists of an inner tracking detector (ID) surrounded by a thin superconducting solenoid providing a \SI{2}{\tesla} axial magnetic field,
electromagnetic and hadron calorimeters, and a muon spectrometer (MS).
The inner tracking detector covers the pseudorapidity range \(|\eta| < 2.5\).
It consists of silicon pixel, silicon microstrip, and transition radiation tracking detectors.
Lead/liquid-argon (LAr) sampling calorimeters provide electromagnetic (EM) energy measurements with high granularity.
A steel/scintillator-tile hadron calorimeter covers the central pseudorapidity range (\(|\eta| < 1.7\)).
The endcap and forward regions are instrumented with LAr calorimeters
for EM and hadronic energy measurements up to \(|\eta| = 4.9\).
The muon spectrometer surrounds the calorimeters and is based on
three large air-core toroidal superconducting magnets with eight coils each.
The field integral of the toroids ranges between \num{2.0} and \SI{6.0}{\tesla\metre}
across most of the detector.
The muon spectrometer includes a system of precision tracking chambers and fast detectors for triggering.
Only runs with stable colliding beams and in which all relevant detector components were functional are used.
 
A two-level trigger system,
with the first level implemented in custom hardware
and followed by a software-based second level,
is used to reduce the trigger rate to around 1~kHz for offline storage~\cite{TRIG-2016-01}.
Events in this analysis were recorded using single-lepton triggers.
To maximise the event selection efficiency,
multiple triggers were used, either with low transverse momentum (\pT)
thresholds and lepton identification and isolation requirements,
or with higher \pT\ thresholds but looser identification criteria and no isolation requirements.
Slightly different sets of triggers were used for 2015 and 2016--2018 data
due to the increase in the average number of $pp$ interactions per bunch
crossing (pile-up).
The minimum \pT\ required by the triggers was increased
to keep both trigger rate and data storage within their limits.
For muons, the lowest \pT\ threshold was 20~(26)~GeV in 2015 (2016--2018),
while for electrons, triggers with a minimum \pT\ threshold of 24~(26)~GeV were used~\cite{TRIG-2018-05}.
Simulated events are also required to satisfy the trigger criteria.
 
Signal and background processes were modelled with Monte Carlo (MC) simulation samples.
The $pp \rightarrow \tbhplus$ process followed by \hplustb\ decay was modelled with \MGMCatNLO~\cite{Degrande:2015vpa}
at next-to-leading order (NLO) in QCD~\cite{Alwall:2014hca}
using a four-flavour scheme (4FS) implementation
with the NNPDF2.3NLO~\cite{Ball:2012cx} parton distribution function (PDF).
Parton showers (PS) and hadronisation were modelled by \pythia~8.212~\cite{Sjostrand:2007gs}
with a set of underlying-event (UE) parameters tuned to ATLAS data and named the A14 tune~\cite{ATL-PHYS-PUB-2014-021}.
Dynamic QCD factorisation and renormalisation scales, \muf\ and \mur,
were set to $\frac{1}{3}\sum_i\sqrt{m(i)^2 + \pT(i)^2}$,
where $i$ runs over the final-state particles (\hplus, $t$ and $b$) used in the generation.
Only the \hplus\ decay into $tb$ is considered.
For the simulation of the $tbH^+$ process the narrow-width approximation was used.
This assumption has negligible impact on the analysis for the models considered in this paper,
as the experimental resolution is much larger than the \hplus\ natural width~\cite{Arhrib_2020}.
Interference with the SM $t\bar t b\bar b$ background is neglected.
A total of 18 \hplus\ mass hypotheses are used,
with 25~\GeV\ mass steps between a \hplus\ mass of 200~\GeV\ and 300~\GeV,
50~\GeV\ steps between 300~\GeV\ and 400~\GeV,
100~\GeV\ steps between 400~\GeV\ and 1000~\GeV, and 200~\GeV\ steps from 1000~\GeV\ to 2000~\GeV.
The step sizes were chosen to match the experimental mass resolution of the \hplus\ signal.
 
The production of \ttbar~+~jets events was modelled using the \POWHEGBOX~\cite{Nason:2004rx,Frixione:2007vw,Alioli:2010xd,Campbell:2014kua}
v2 generator in the five-flavour scheme (5FS),
which provides matrix elements (ME) at NLO in QCD,
with the NNPDF3.0NLO PDF set~\cite{Ball:2014uwa}.
The $h_{\mathrm{damp}}$ parameter,
which controls the transverse momentum of the first additional emission beyond the Born configuration,
was set to 1.5$m_t$~\cite{ATL-PHYS-PUB-2016-020},
where $m_t$ is the mass of the top quark.
Parton showers and hadronisation were modelled by \pythia~8.230~\cite{Sjostrand:2014zea} with the A14 tune for the UE.
The scales \muf\ and \mur\ were set to the default scale $\sqrt{m_t^2+ p_{\mathrm{T},t}^2}$.
The sample was normalised to the \toppp~2.0~\cite{Czakon:2011xx}
theoretical cross-section of $832^{+46}_{-51}$~pb,
calculated at next-to-next-to-leading order (NNLO) in QCD including resummation of
next-to-next-to-leading logarithmic (NNLL) soft gluon terms~\cite{Cacciari:2011hy,Baernreuther:2012ws,Czakon:2012zr,Czakon:2013goa}.
The generation of the \ttbar\ + jets events was performed both inclusively of additional jet flavour,
and also with dedicated filtered samples,
requiring $b$- or $c$-hadrons in addition to those arising from the decays of the top quarks.
Events generated with no extra $b$-hadrons were taken from the unfiltered sample and merged with the \ttbar\ + jets events from the filtered sample,
taking the appropriate cross-section and filter efficiencies into account.
 
Single-top $t$-channel production was modelled using the \POWHEGBOXV{v2} generator at NLO in QCD,
generated in the 4FS with the NNPDF3.0NLOnf4 PDF set~\cite{Ball:2014uwa}.
The scales \muf\ and \mur\ were set to
$\sqrt{m_b^2+ p_{\mathrm{T},b}^2}$ following the recommendation in Ref.~\cite{Frederix:2012dh}.
Single-top $tW$ and $s$-channel production was modelled using the \POWHEGBOXV{v2} generator at NLO in QCD,
generated in the 5FS with the NNPDF3.0NLO PDF set.
The scales \muf\ and \mur\ were set to the default scale,
which is equal to the top-quark mass.
For $tW$ associated production,
the diagram removal scheme~\cite{Frixione:2008yi} was employed to handle the interference with \ttbar\ production~\cite{ATL-PHYS-PUB-2016-020}.
All single-top events were showered with \pythia~8.230.
 
Production of vector bosons with additional jets was simulated with the \sherpa~2.2.1 generator~\cite{Bothmann:2019yzt}.
Matrix elements with NLO accuracy for up to two partons, and with leading-order (LO) accuracy for up to four partons,
were calculated with the Comix~\cite{Gleisberg:2008fv} and OpenLoops~\cite{Cascioli:2011va,Denner:2016kdg} libraries.
The default \sherpa\ PS algorithm~\cite{Schumann:2007mg} based on Catani--Seymour dipole factorisation and
the cluster hadronisation model~\cite{Winter:2003tt} was used.
It employs the dedicated set of tuned parameters developed by the \sherpa\ authors for this version,
based on the NNPDF3.0NNLO PDF set.
The NLO ME of a given jet multiplicity were matched to the PS using a colour-exact variant of the MC@NLO algorithm~\cite{Hoeche:2011fd}.
Different jet multiplicities were then merged into an inclusive sample using an improved CKKW matching procedure~\cite{Catani:2001cc,Hoeche:2009rj},
which is extended to NLO accuracy using the MEPS@NLO prescription~\cite{Hoeche:2012yf}.
The merging cut was set to 20~\GeV.
 
The production of $t\bar t V$ events, i.e.\ $t\bar t W$ or $t\bar t Z$,
was modelled using the \MGMCatNLOV{2.3.3} generator,
which provides ME at NLO in QCD with the NNPDF3.0NLO PDF set.
The scales \muf\ and \mur\ were set to
the default scale $\frac{1}{2}\sum_i\sqrt{m_i^2+{\pT}_i^2}$,
where the sum runs over all the particles generated in the ME calculation.
The events were showered with \pythia~8.210.
Additional $t\bar t V$ samples were produced with the \sherpa~2.2.0~\cite{Bothmann:2019yzt} generator at LO accuracy,
using the MEPS@LO prescription with up to one additional parton for the $t\bar t Z$ sample
and two additional partons for $t\bar t W$.
A dynamic scale \mur\ is used,
defined similarly to that of the nominal \MGMCatNLO samples.
The CKKW matching scale of the additional emissions was set to 30~\GeV.
The default \sherpa~2.2.0 PS was used along with the NNPDF3.0NNLO PDF set.
The production of $t\bar t H$ events was modelled in the 5FS using the \POWHEGBOX~\cite{Hartanto:2015uka}
generator at NLO with the NNPDF3.0NLO PDF set.
The $h_{\mathrm{damp}}$ parameter was set to $\frac{3}{4} (2m_t + m_H ) = 352.5$~\GeV,
and the events are showered with \pythia~8.230.
 
Diboson ($VV$) samples were simulated with the \sherpa~2.2 generator.
Multiple ME calculations were matched and merged with the \sherpa\ PS
using the MEPS@NLO prescription.
For semileptonically and fully leptonically decaying diboson samples,
as well as loop-induced diboson samples,
the virtual QCD correction for ME at NLO accuracy were provided by the OpenLoops library.
For electroweak $VVjj$ production,
the calculation was performed in the $G_{\mu}$-scheme~\cite{Denner:2003iy},
ensuring an optimal description of pure electroweak interactions at the electroweak scale.
All samples were generated using the NNPDF3.0NNLO PDF set,
along with the dedicated set of tuned PS parameters developed by the \sherpa\ authors.
 
Other minor backgrounds ($tHjb$, $tHW$, $tZq$, $tZW$ and four top quarks) were also simulated and accounted for,
even though they contribute less than 1\% in any analysis region.
All samples and their basic generation parameters are summarised in Table~\ref{tab:MCsummary}.
 
\begin{table}[htb]
\caption{Nominal simulated signal and background event samples.
The ME generator,
PS generator and calculation accuracy of the
cross-section in QCD used for normalisation
(aNNLO stands for approximate NNLO in QCD)
are shown together with the applied PDF set.
Either \sherpa~2.2.1 or \sherpa~2.2.2 was used for different diboson contributions.
The rightmost column shows whether fast or full simulation was used to produce the samples.}
\small
\begin{center}
\begin{tabular}{llllll}
\hline
Physics process    &  ME generator       & PS generator     & Normalisation &  PDF set   & Simulation \\
\hline \hline
$tbH^+$            &  \madgraphShort\ 2.6.2  &  \pythia\ 8.212   &   NLO   &    NNPDF2.3NLO  &  Fast \\
\hline
$t\bar t$ + jets   &  \POWHEGBOXV{v2}       &  \pythia\ 8.230   &   NNLO+NNLL  &    NNPDF3.0NLO  &  Fast  \\
\hline
Single-top $t$-chan  &  \POWHEGBOXV{v2}       &  \pythia\ 8.230   &   aNNLO      &    NNPDF3.0NLOnf4  &  Full  \\
Single-top $tW$    &  \POWHEGBOXV{v2}       &  \pythia\ 8.230   &   aNNLO      &    NNPDF3.0NLO  &  Full \\
Single-top $s$-chan  &  \POWHEGBOXV{v2}       &  \pythia\ 8.230   &   aNNLO      &    NNPDF3.0NLO  &  Full \\
\hline
$V$ + jets         &  \sherpa\ 2.2.1    &  \sherpa\ 2.2.1   &  NNLO        &    NNPDF3.0NNLO  & Full  \\
\hline
$t\bar t V$        &  \madgraphShort\ 2.3.3  &  \pythia\ 8.210   &   NLO   &    NNPDF3.0NLO   & Full \\
$t\bar t H$        &  \POWHEGBOXV{v2}       &  \pythia\ 8.230   &   NLO        &    NNPDF3.0NLO   & Full  \\
\hline
Diboson            &  \sherpa\ 2.2      &  \sherpa\  2.2    &  NLO         &    NNPDF3.0NNLO  & Full  \\
\hline
$tHjb$             &  \madgraphShort\ 2.6.0  &  \pythia\ 8.230   &   NLO   &    NNPDF3.0NLOnf4  &  Full \\
$tHW$              &  \madgraphShort\ 2.6.2  &  \pythia\ 8.235   &   NLO   &    NNPDF3.0NLO  &   Full \\
$tZq$              &  \madgraphShort\ 2.3.3  &  \pythia\ 8.212   &   NLO   &    CTEQ6L1LO    &   Full  \\
$tZW$              &  \madgraphShort\ 2.3.3  &  \pythia\ 8.212   &   NLO   &    NNPDF3.0NLO  &   Full  \\
Four top quarks     &  \madgraphShort\ 2.3.3  &  \pythia\ 8.230   &   NLO   &    NNPDF3.1NLO  &   Fast  \\
\hline
\end{tabular}
\end{center}
\label{tab:MCsummary}
\end{table}
 
Most of the samples mentioned above were produced using
the full ATLAS detector simulation~\cite{SOFT-2010-01}
based on $\GEANT4$~\cite{Agostinelli:2002hh},
and the rest were produced using fast simulation~\cite{ATL-PHYS-PUB-2010-013},
where the complete $\GEANT4$ simulation of the calorimeter response is replaced by a
detailed parameterisation of the shower shapes,
as shown in Table~\ref{tab:MCsummary}.
For the observables used in this analysis,
the two simulations were found to give compatible results.
Additional pile-up interactions,
simulated with \pythia~8.186 using the A3 set of tuned parameters~\cite{ATL-PHYS-PUB-2016-020},
were overlaid onto the simulated hard-scatter event.
All simulation samples were reweighted such that the distribution of the number of pile-up interactions matches that of the data.
In all samples the top-quark mass was set to 172.5~\GeV,
and the decays of $b$- and $c$-hadrons were performed by EvtGen v1.2.0~\cite{Lange:2001uf},
except in samples simulated by the \sherpa\ event generator.

 
\section{Object reconstruction and event selection}
\label{sec:objects}
Charged leptons and jets,
including those compatible with the hadronisation of $b$-quarks,
are the main reconstructed objects used in this analysis.
Electrons are reconstructed from energy clusters in the electromagnetic calorimeter associated with tracks reconstructed in the ID~\cite{EGAM-2018-01},
and are required to have $|\eta| < 2.47$.
Candidates in the calorimeter transition region ($1.37<| \eta |<1.52$) are excluded.
Electrons must satisfy the \textit{tight} identification criterion
described in Ref.~\cite{ATLAS-CONF-2016-024},
based on shower-shape and track-matching variables.
Muons are reconstructed from either track segments or full tracks in the MS which are
matched to tracks in the ID.
Tracks are then re-fit using information from both detector systems.
Muons must satisfy the \textit{medium} identification criterion~\cite{PERF-2015-10}.
Muons are required to have $|\eta| < 2.5$.
To reduce the contribution of leptons from hadronic decays (non-prompt leptons),
both the electrons and muons must satisfy isolation criteria.
These criteria include both track and calorimeter information,
and have an efficiency of 90\% for leptons with a \pT\ greater than 25~\GeV,
rising to 99\% above 60~\GeV,
as measured in $Z\rightarrow ee$ and $Z\rightarrow \mu\mu$ data samples~\cite{EGAM-2018-01,PERF-2015-10}.
Finally, the lepton tracks must point to the
primary vertex of the event,\footnote{Events are required to have at least
one reconstructed vertex with three or more associated tracks which have $\pT> 400$~\MeV.
The primary vertex is chosen as the vertex candidate with the largest sum of
the squared transverse momenta of associated tracks.}
the longitudinal impact parameter must satisfy $|z_0| < 0.5$~mm and
the transverse impact parameter significance must satisfy $|d_0|/\sigma_{d_0}<5\, (3)$ for electrons (muons).
 
Jets are reconstructed from three-dimensional topological energy clusters~\cite{PERF-2014-07} in the calorimeter using the anti-$k_t$ jet algorithm~\cite{Cacciari:2008gp} with a radius parameter of 0.4.
Each topological cluster is calibrated to the electromagnetic scale response prior to jet reconstruction.
The reconstructed jets are then calibrated with a series of simulation-based corrections and
in situ techniques based on 13~\TeV\ data~\cite{PERF-2016-04}.
After energy calibration, jets are required to have $\pT >25$~\GeV\ and $|\eta|<2.5$.
Quality criteria are imposed to identify jets arising from non-collision sources or detector noise,
and any event containing such a jet is removed~\cite{ATLAS-CONF-2015-029}.
Finally, to reduce the effect of pile-up,
jets with $\pT <120$~\GeV\ and $|\eta|<2.4$ are matched to tracks with $\pT >0.5$~\GeV,
thus ensuring they originate from the primary vertex.
This algorithm is known as the jet vertex tagger (JVT)~\cite{PERF-2014-03}.
To identify jets containing $b$-hadrons,
referred to as $b$-jets in the following,
the MV2c10 tagger algorithm~\cite{PERF-2016-05},
which combines impact parameter information with the explicit identification
of secondary and tertiary vertices within the jet into a multivariate discriminant,
is used.
Jets are $b$-tagged by requiring the discriminant output to be above a threshold,
providing a specific $b$-jet efficiency in simulated $t\bar t$ events.
A criterion with an efficiency of 70\% is used to determine
the $b$-jet multiplicity in this analysis.
For this working point and for the same \ttbar\ sample,
the $c$-jet and light-flavour-quark or gluon jet (light-jets) rejection factors are 8.9 and 300,
respectively~\cite{FTAG-2018-01}.
 
To avoid counting a single detector signal as more than one lepton or jet,
an overlap removal procedure is applied.
First,
the closest jet within
$\Delta R_y=\sqrt{(\Delta y)^2+(\Delta\phi)^2}=0.2$ of a selected electron is removed.\footnote{The rapidity
is defined as $y=\frac{1}{2} \ln\frac{E+p_z}{E-p_z}$,
where $E$ is the energy and $p_z$ is the momentum component along the beam pipe.}
If the nearest jet surviving that selection is within $\Delta R_y=0.4$ of the electron,
the electron is discarded.
Muons are removed if they are separated from the nearest jet by $\Delta R_y<0.4$,
which reduces the background from semileptonic decays of heavy-flavour hadrons. 
However, if this jet has fewer than three associated tracks, the muon is kept and the jet is removed instead;
this avoids an inefficiency for high-energy muons undergoing significant energy loss in the calorimeter.
 
The missing transverse momentum (of size \MET)
in the event is computed as the negative vector sum of the
\pT\ of all the selected electrons, muons and jets described above,
with a correction for soft energy not associated with any of the hard objects.
This additional `soft term' is calculated from ID tracks matched to the primary vertex to make it resilient to pile-up contamination~\cite{ATL-PHYS-PUB-2015-027}.
The missing transverse momentum is not used in the event selection,
but included in the multivariate discriminant used in the analysis.
 
Events are required to have exactly one electron or muon, with $\pT > 27$~\GeV,
within $\Delta R<0.15$ of a lepton of the same flavour reconstructed by the trigger algorithm,
and at least five jets,
at least three of which must be $b$-tagged.
The total event acceptance for the \hplus\ signal samples ranges from
2\% (at 200~\GeV) to 8.5\% (at 1000~\GeV).
Above 1000~\GeV,
the acceptance decreases due to the boosted topology of the events,
which fail the requirement on jet multiplicity.
At 2000~\GeV, the acceptance is 6\%.
The selected events are categorised into four separate regions
according to the number of reconstructed jets (j) and $b$-jets (b) in the event,
in order to improve the sensitivity of the fit and constrain some of the systematic uncertainties.
The analysis regions are 5j3b, 5j$\ge$4b, $\ge$6j3b and $\ge$6j$\ge$4b,
where XjYb means that X jets are found in the event, and among them Y are $b$-tagged.
In addition, the $\ge$5j2b region is used to derive data-based corrections,
which are implemented to improve the level of agreement between simulation and data.

 
\section{Background modelling}
\label{sec:background}
With the isolation criteria applied both at the trigger and analysis level,
as well as the purity-enhancing identification criteria used for
electrons and muons (Section~\ref{sec:objects}),
the background due to non-prompt leptons is expected to be negligible.
To confirm this assumption,
the ratio  $(N_{\mathrm{Data}} -N^{\mathrm{total}}_{\mathrm{MC}} ) / N^{\mathrm{total}}_{\mathrm{MC}}$
was checked and found to not decrease when moving from a loose to a tight isolation selection.
Such behaviour shows that the non-prompt-lepton background,
which is not present in the simulation,
provides a negligible contribution to the data,
as expected given that non-prompt leptons are unlikely to be isolated in data.
If data and the MC predictions differed due to a mismodelling of the other backgrounds,
tighter isolation requirements would remove events in data and MC simulation alike.
All backgrounds in this analysis are estimated using the simulation samples described in Section~\ref{sec:dataMC}.
 
To define the background categories in the likelihood fit (Section~\ref{sec:results}),
the \ttbar~+~jets background is categorised according to the flavour of
the jets in the event.
Generator-level particle jets are reconstructed from stable particles
(mean lifetime $\tau > 3 \times 10^{-11}$~s)
using the anti-$k_t$ algorithm with a radius parameter $R$ = 0.4,
and are required to have \pT > 15~GeV and $|\eta|$ < 2.5.
The flavour of a jet is determined by counting the number of $b$- or $c$-hadrons within
$\Delta R =$ 0.4 of the jet axis.
Jets matched to one or more $b$-hadrons,
of which at least one must have \pT\ above 5~GeV,
are labelled as $b$-jets;
$c$-jets are defined analogously,
only considering jets not already defined as $b$-jets.
Events that have at least one $b$-jet,
not including heavy-flavour jets from top-quark or $W$-boson decays,
are labelled as \ttbin;
those with no $b$-jets but at least one $c$-jet are labelled as \ttcin.
Finally,
events not containing any heavy-flavour jets,
aside from those from top-quark or $W$-boson decays,
are labelled as \ttlin.
 
After the event selection, \ttbar~+~jets constitutes the main background.
It is observed that the simulation of \ttbar~+~jets does not properly model
high jet multiplicities nor the hardness of additional jet emissions,
and data-based corrections are applied to the simulation~\cite{ATL-PHYS-PUB-2018-009, collaboration2020measurements}.
Given that the additional jets in the \ttbar~+~jets sample are simulated in the parton shower,
the mentioned mismodelling is expected to be independent of whether the additional jets are $b$-tagged or not.
Therefore, data and MC predictions are compared and reweighting factors are derived
in a sample with at least five jets and exactly two $b$-tagged jets,
and then applied in the 3b and $\ge$4b regions.
The level of agreement between data and simulation
in these regions improves to the point where the remaining differences
are well within the model's systematic uncertainty.
The reweighting factors are expressed as:
\begin{equation*}
R(x) = \frac{ N_{\mathrm{Data}}(x)-N_{\mathrm{MC}}^{\textrm{non-}\ttbar}(x) }
{ N_{\mathrm{MC}}^{\ttbar}(x) },
\end{equation*}
 
\noindent where $x$ is the variable mismodelled by the simulation.
In this context,
\ttlin, \ttbin\ and \ttcin,
as well as $Wt$ single-top contributions,
are included in the \ttbar\ sample.
For the range of \hplus\ masses considered in this analysis and assuming the observed upper limits on the cross-section times branching ratio published in Ref.~\cite{HIGG-2017-04},
signal events contribute less than 1\% to the $\ge$5j2b region and are neglected.
Weights are calculated from the number of jets distribution ($R(\mathrm{nJets})$) first
and subsequently applied to the \HTall\
distributions\footnote{\HTall\ is defined as the scalar sum of the transverse momenta of all jets and the lepton in the event.}
to derive the reweighting factors in the 5j2b, 6j2b, 7j2b and $\ge$8j2b regions ($R(\HTall)$).
Thus,
events are weighted by the product $R(\mathrm{nJets}) \times R(\HTall)$
depending on their jet multiplicity and \HTall\ value.
Figure~\ref{fig:weights} shows $R(\mathrm{nJets})$
in the $\ge$5j2b region and $R(\HTall)$ in the 5j2b, 6j2b, 7j2b and $\ge$8j2b regions.
Among various functions tried, a hyperbola plus a sigmoid functional form
was found to be the best fit to the \HTall\ weight distributions.
 
\begin{figure}[htb]
\centering
\subfloat[$\ge$5j2b]{\includegraphics[width = 0.3\textwidth]{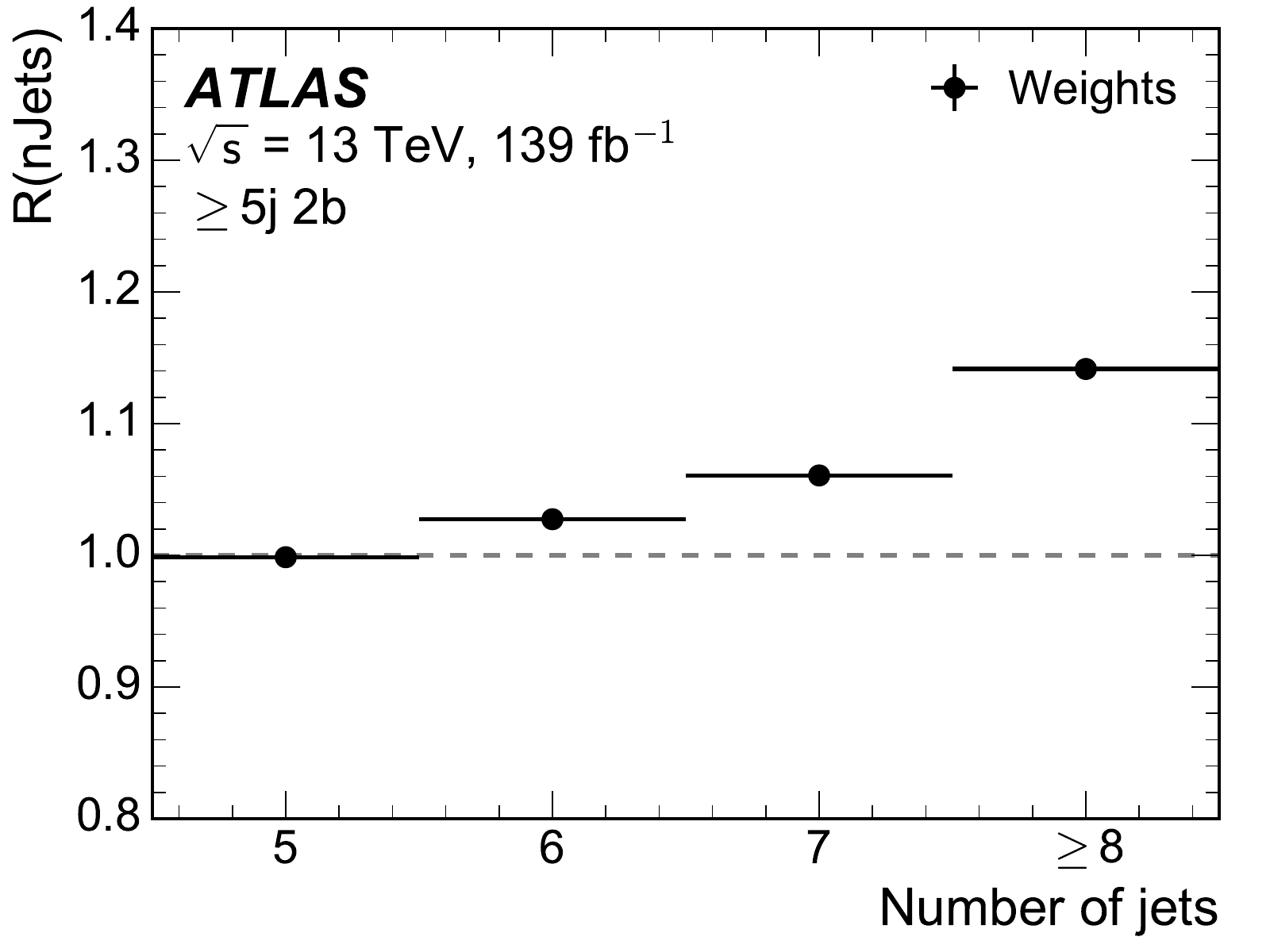}}
\subfloat[5j2b]{\includegraphics[width = 0.3\textwidth]{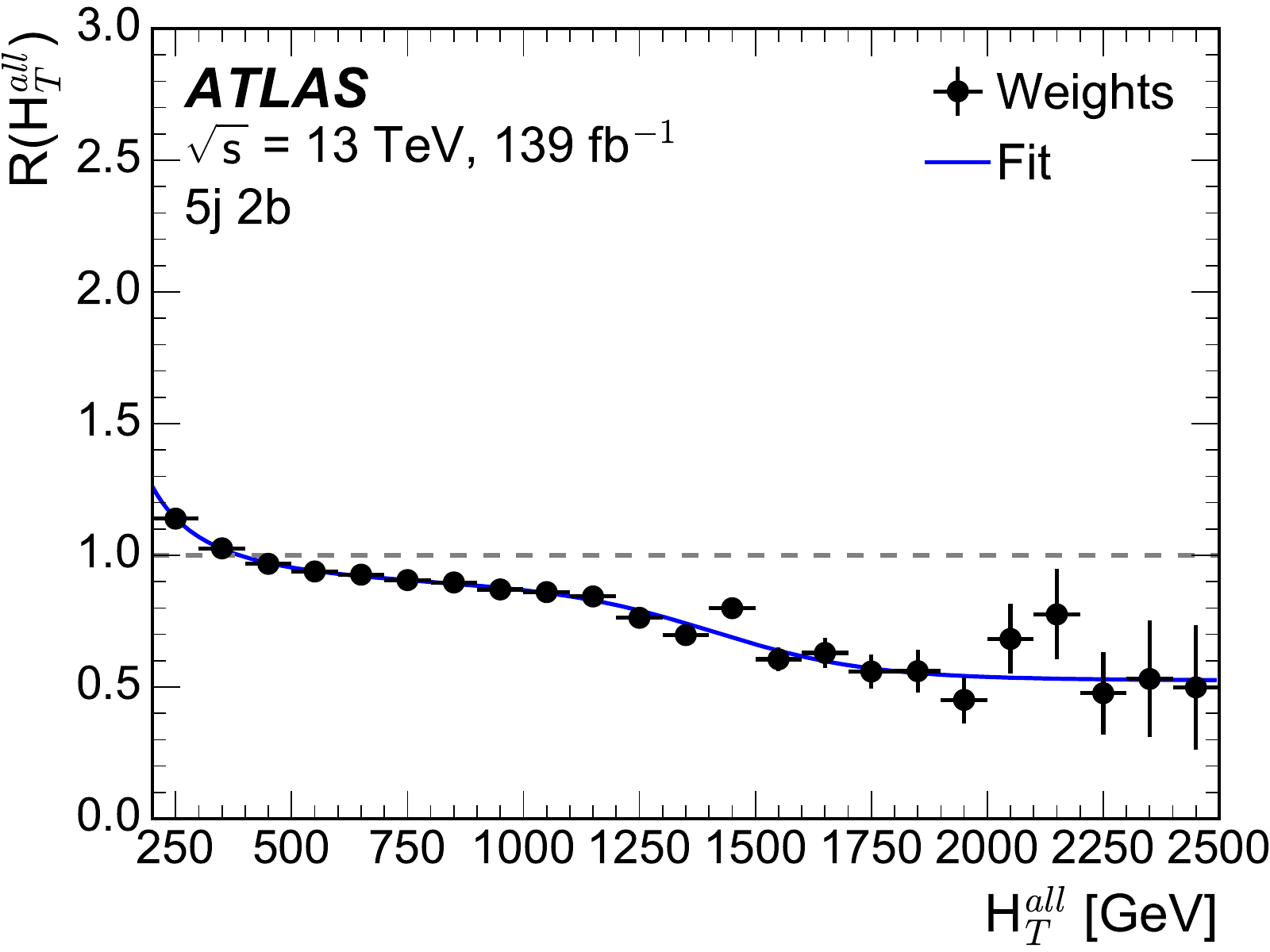}}
\subfloat[6j2b]{\includegraphics[width = 0.3\textwidth]{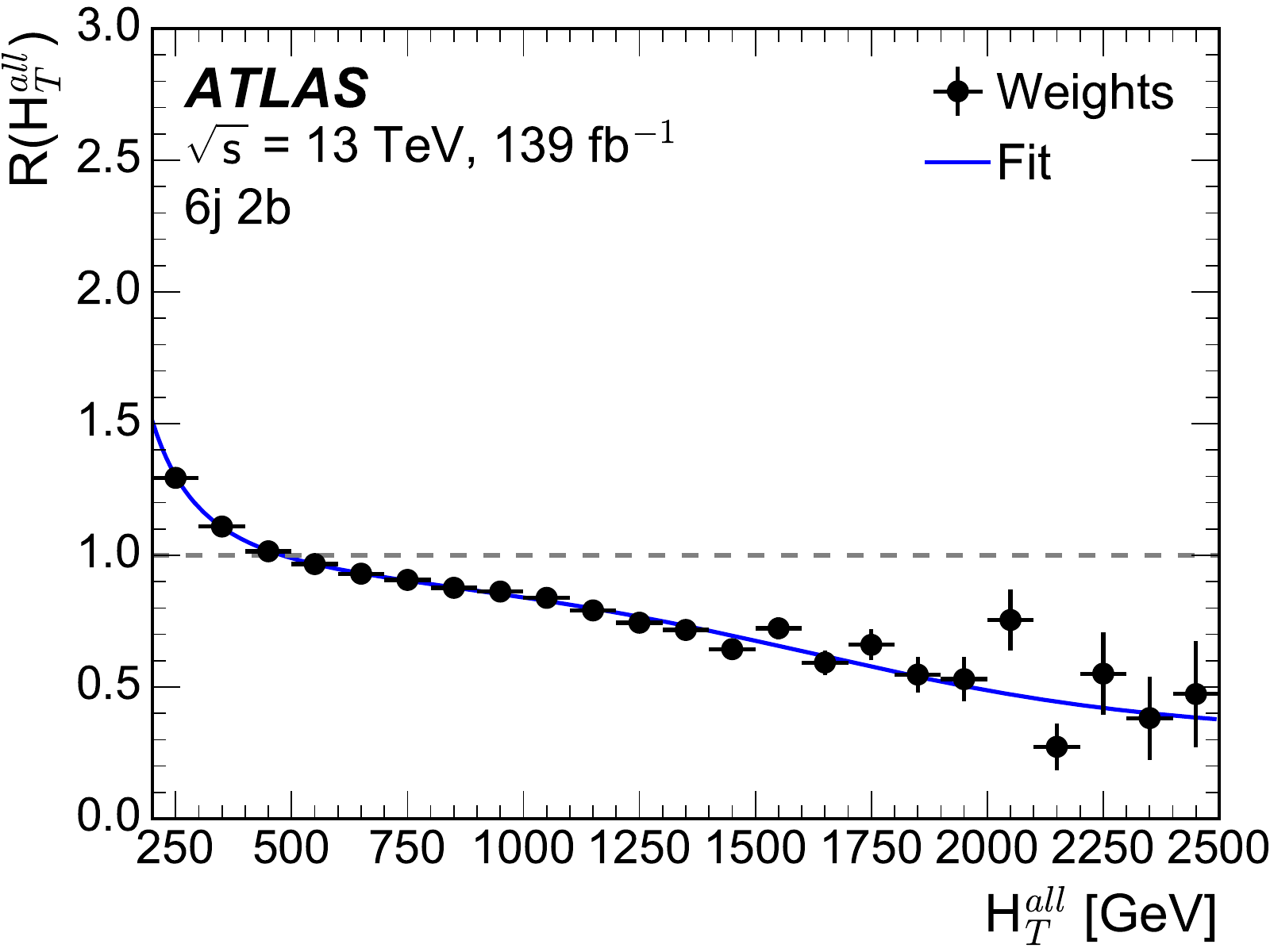}}  \\
\subfloat[7j2b]{\includegraphics[width = 0.3\textwidth]{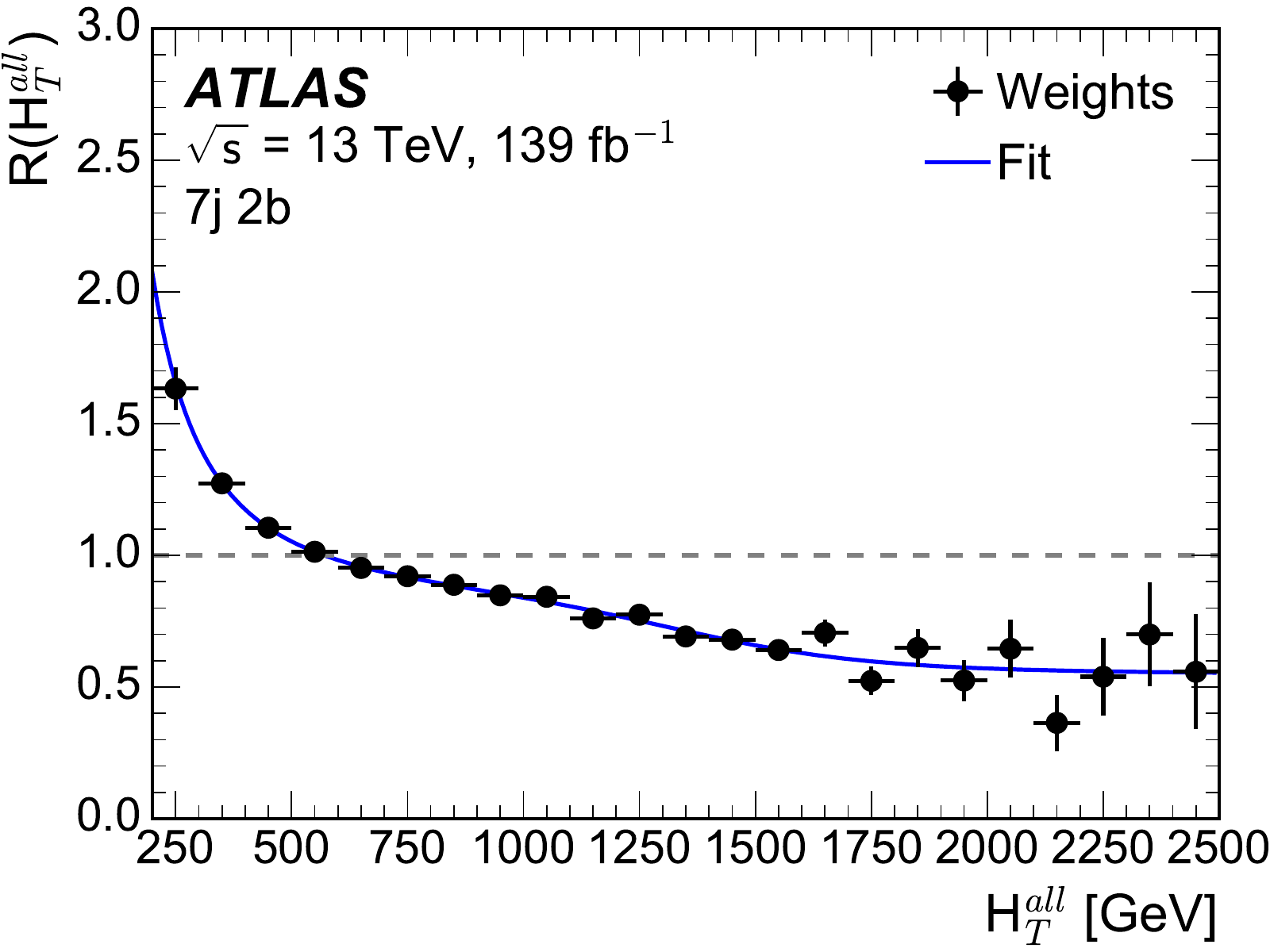}}
\subfloat[$\ge$8j2b]{\includegraphics[width = 0.3\textwidth]{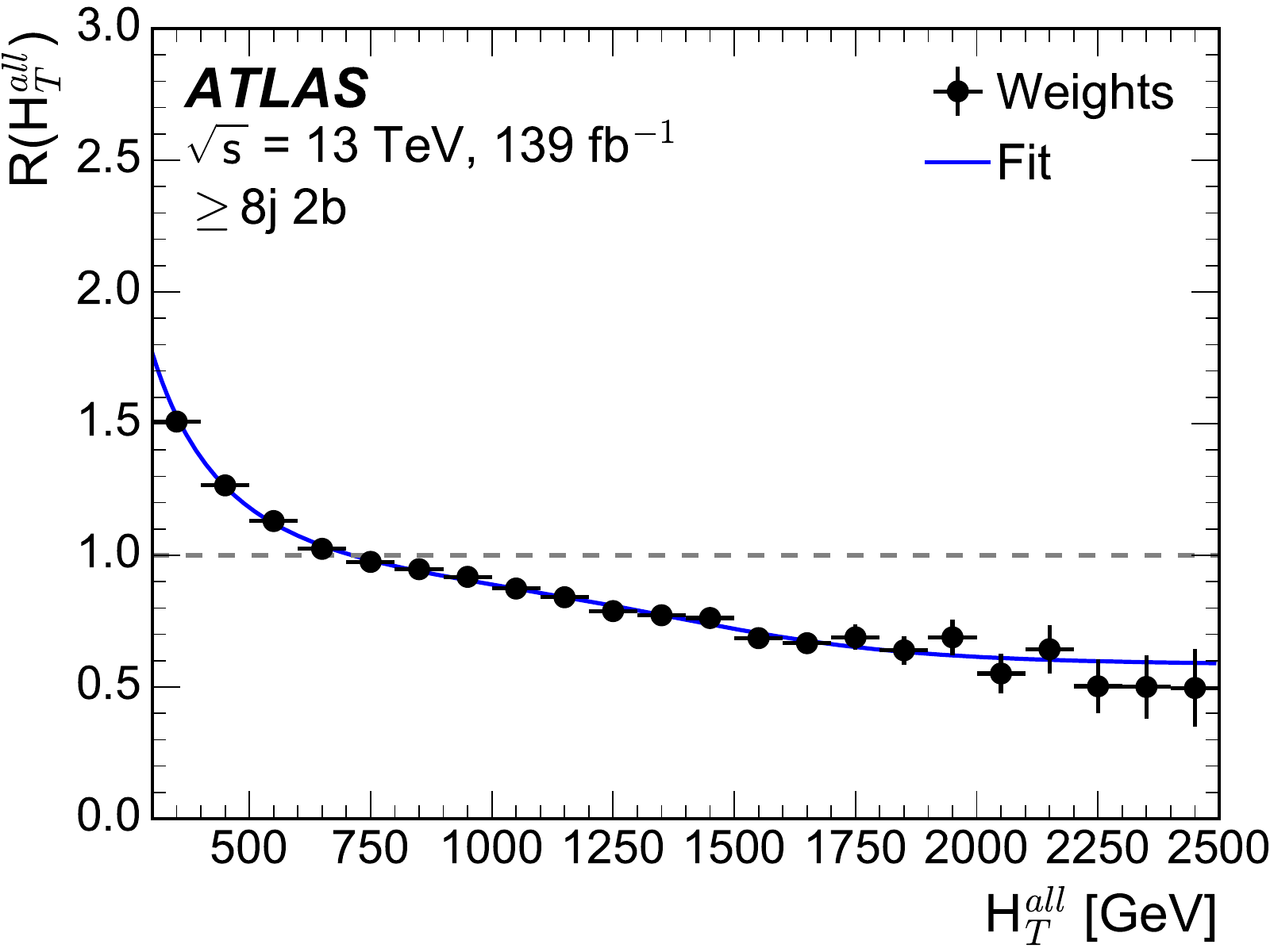}} \\
\caption{Reweighting factors (weights) obtained from the comparison between data and simulation of the number of jets (a)
and \HTall\ for various jet multiplicity selections (b) to (e).
The errors in the data points include the statistical uncertainties in data and MC predictions.}
\label{fig:weights}
\end{figure}
 
After the reweighting, agreement between simulation and data in the analysis regions improves,
as can be seen,
for example,
in Figure~\ref{fig:leadJetReweight},
which shows the leading jet's \pT\ distribution before the fit,
both before and after applying the reweighting.
The final \ttbin\ and \ttcin\ normalisation factors and their uncertainties,
which account for the remaining mismodelling observed after applying the reweighting,
are not applied.
These normalisations are extracted from the fit to data,
as described in Section~\ref{sec:results}.

\begin{figure}[htb]
\centering
\subfloat[5j3b, unweighted]
{\includegraphics[width = 0.249\textwidth]{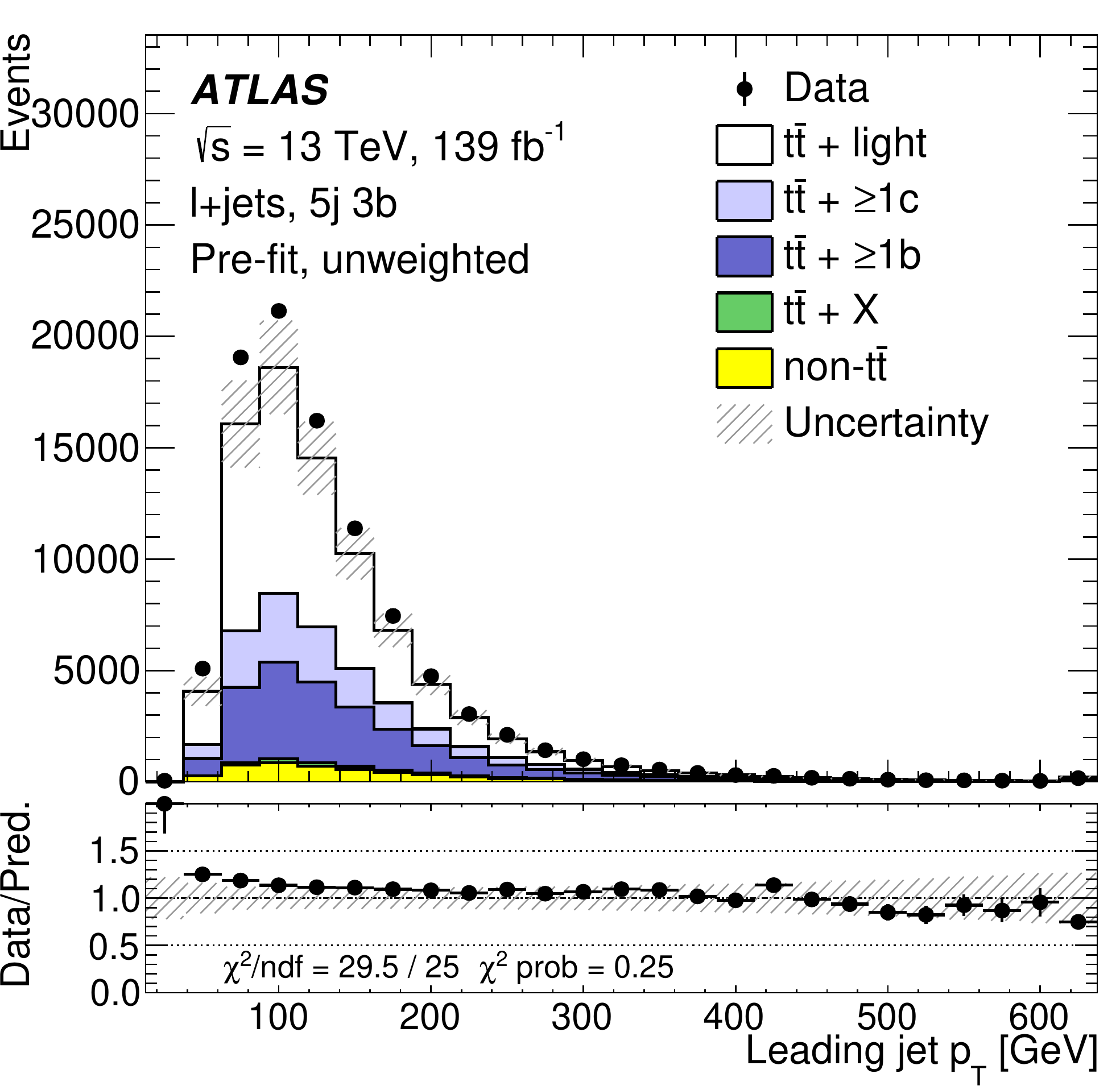}}
\subfloat[5j$\ge$4b, unweighted]
{\includegraphics[width = 0.249\textwidth]{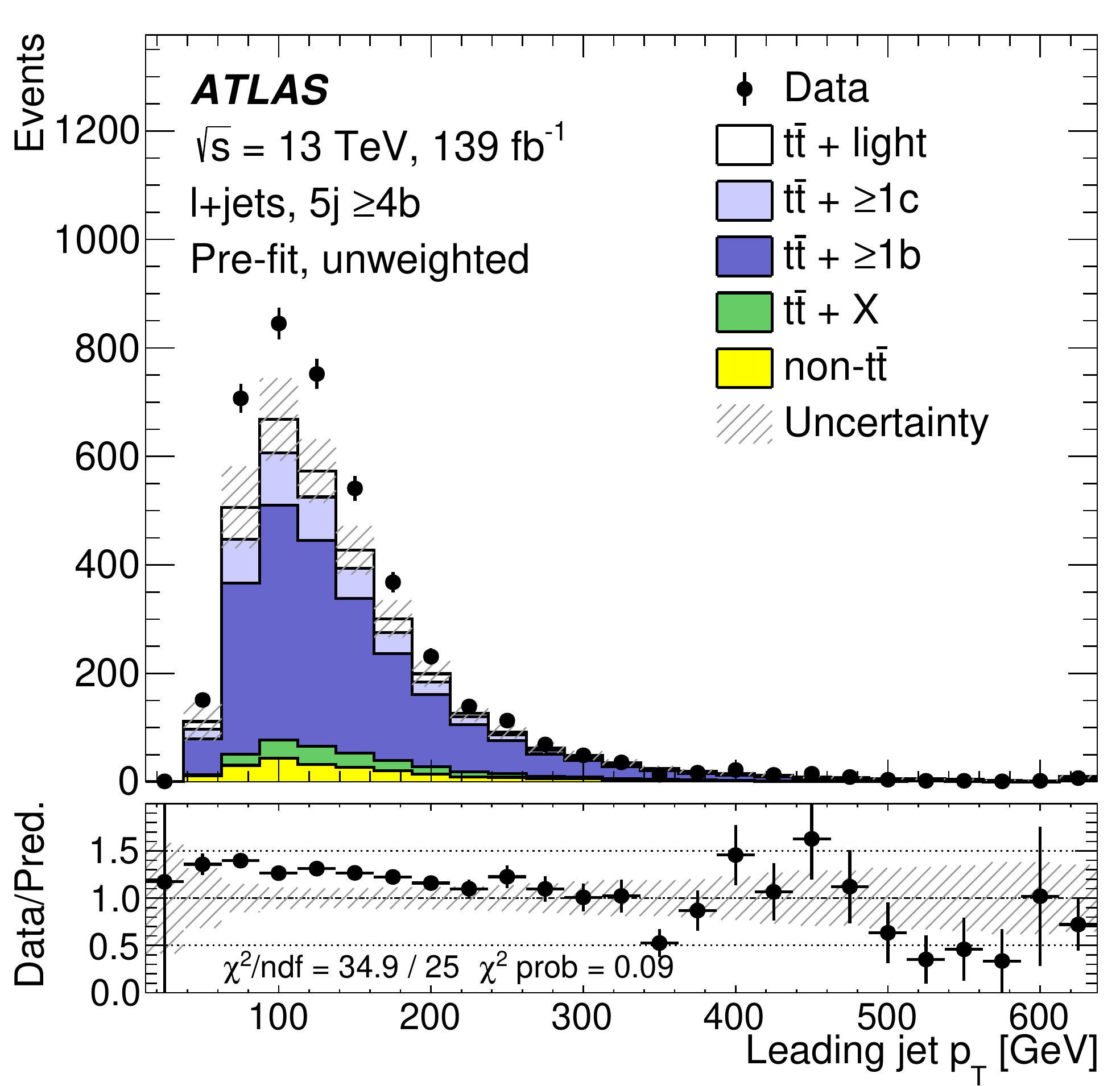}}
\subfloat[$\ge$6j3b, unweighted]
{\includegraphics[width = 0.249\textwidth]{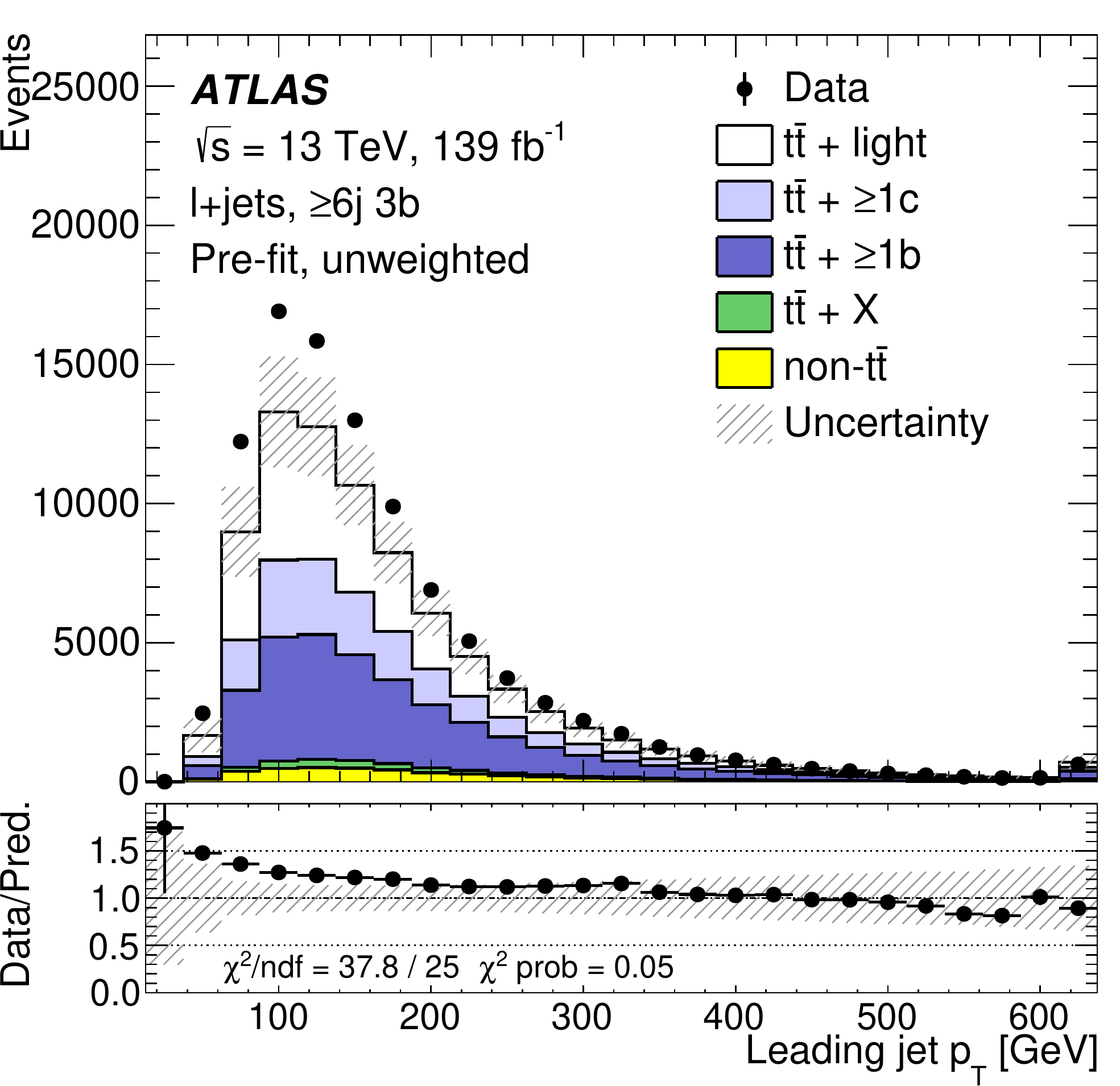}}
\subfloat[$\ge$6j$\ge$4b, unweighted]
{\includegraphics[width = 0.249\textwidth]{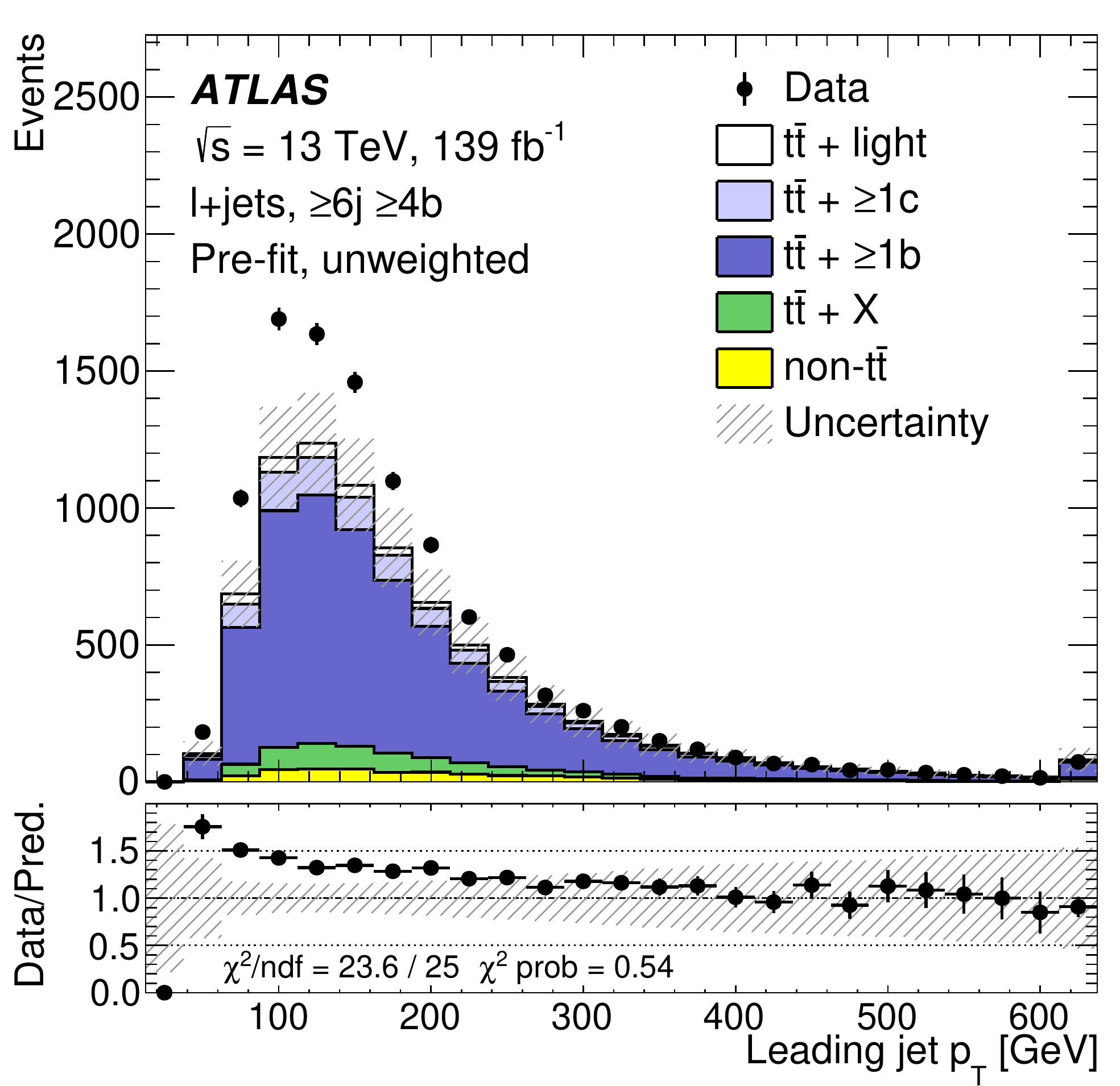}}  \\
\subfloat[5j3b, reweighted]
{\includegraphics[width = 0.249\textwidth]{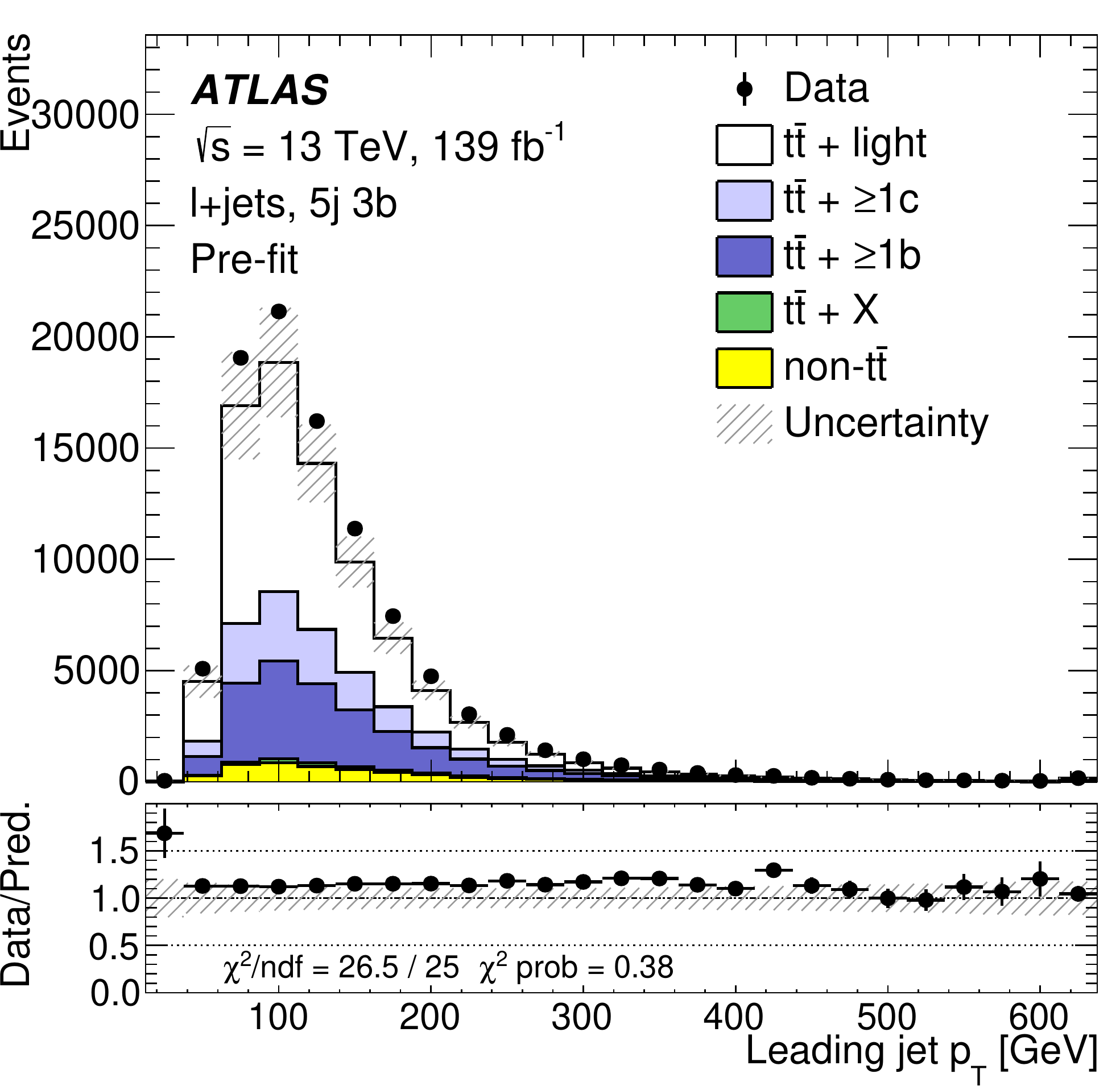}}
\subfloat[5j$\ge$4b, reweighted]
{\includegraphics[width = 0.249\textwidth]{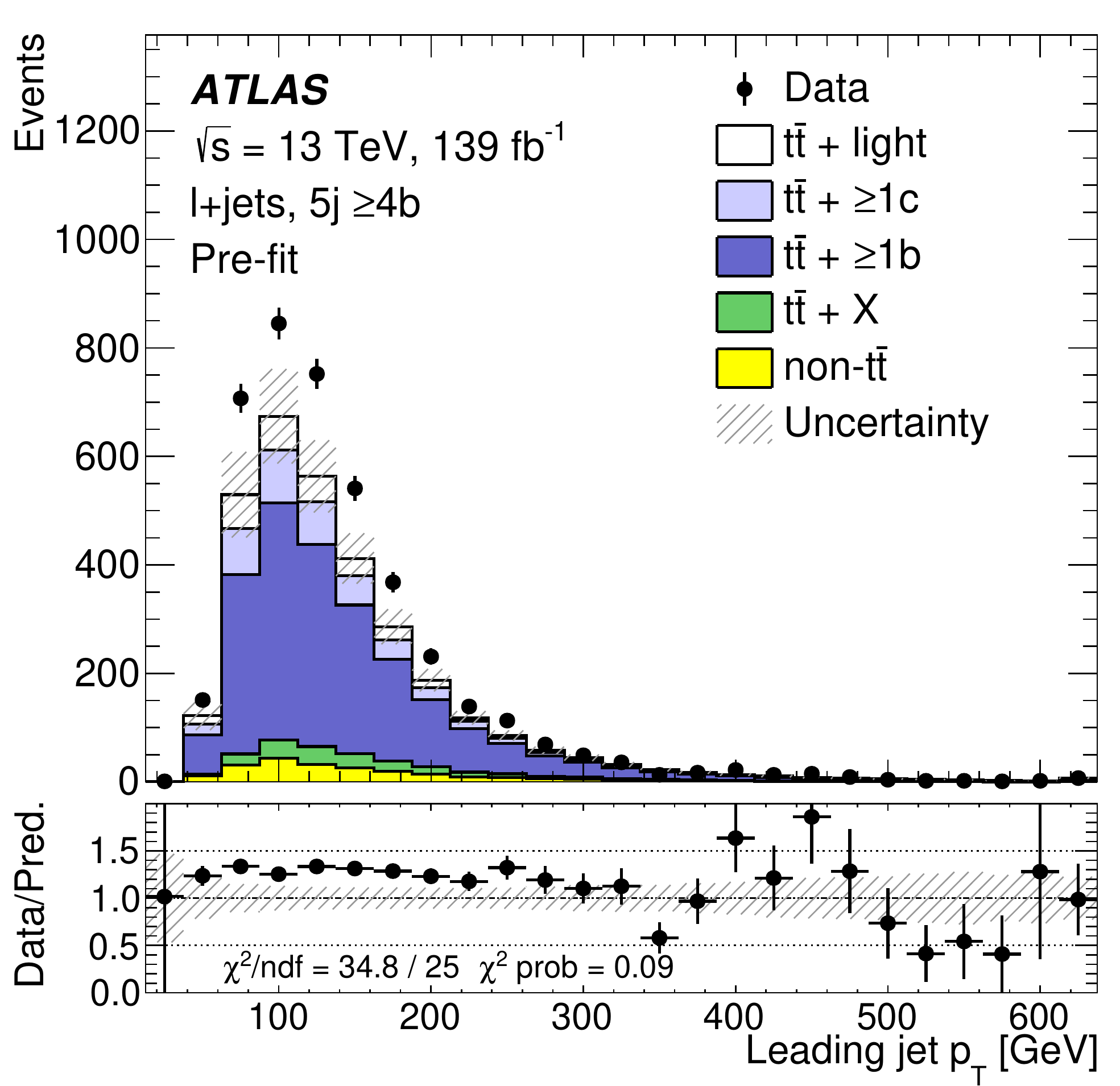}}
\subfloat[$\ge$6j3b, reweighted]
{\includegraphics[width = 0.249\textwidth]{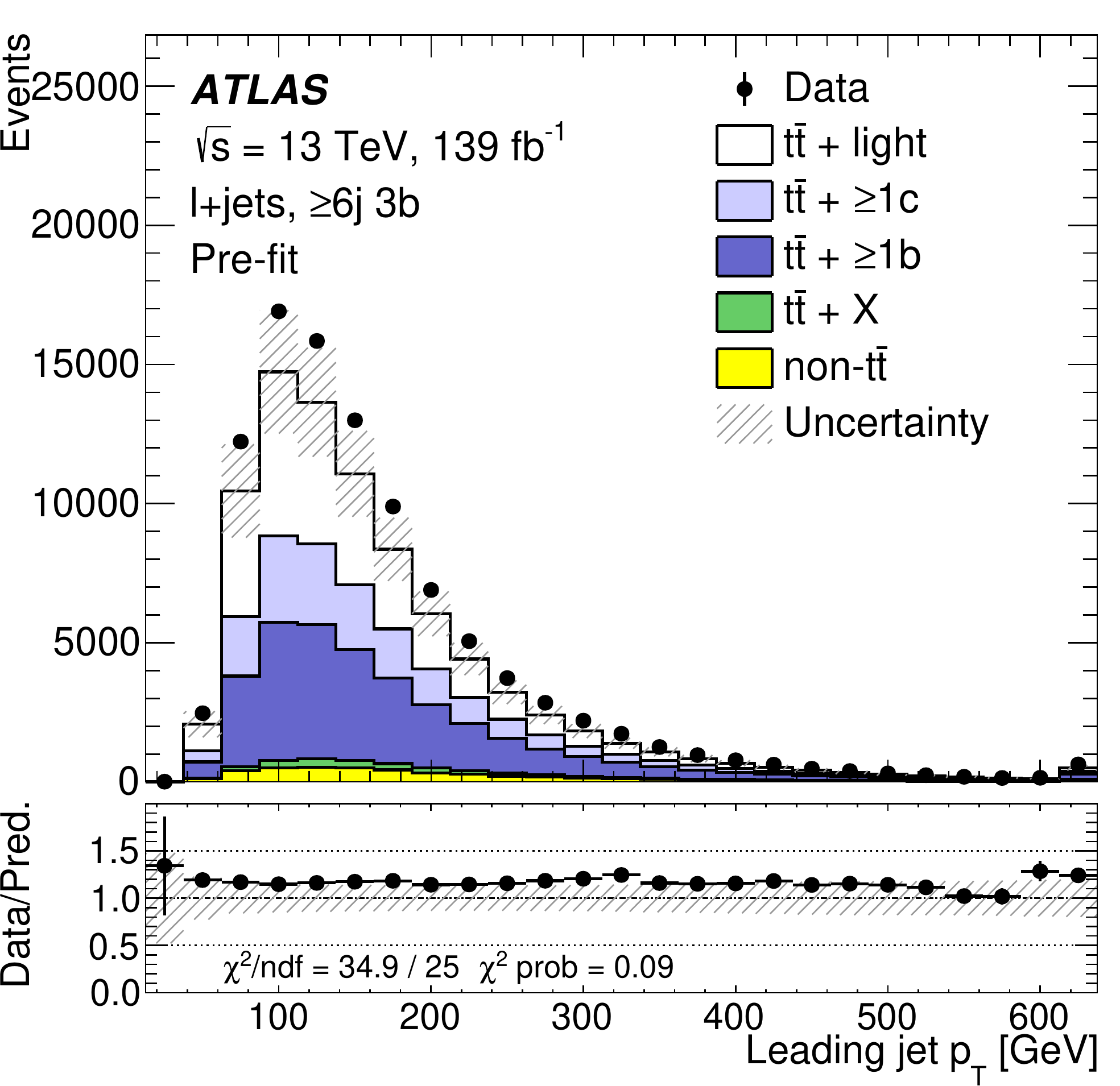}}
\subfloat[$\ge$6j$\ge$4b, reweighted]
{\includegraphics[width = 0.249\textwidth]{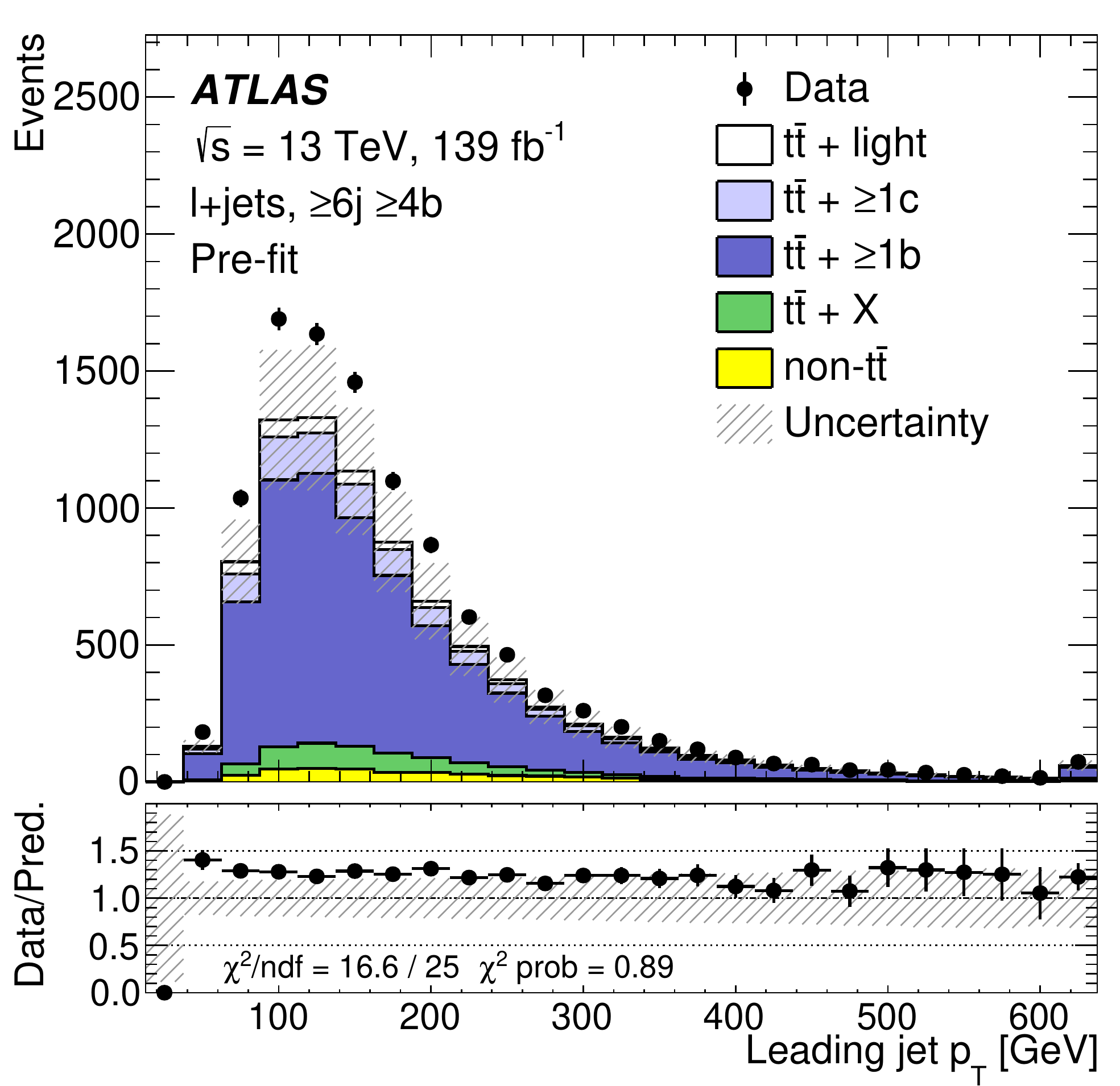}} \\
\caption{Comparison of the predicted leading jet \pT\ and data before the fit in the four analysis regions before (top) and after (bottom) the reweighting was applied.
The uncertainty bands include both the statistical and systematic uncertainties.
Since the normalisations of the \ttbin\ and \ttcin\ backgrounds are allowed to vary in the fit,
no cross-section uncertainties associated with these processes are included.
The lower panels display the ratio of the data to the total prediction.
The hatched bands show the uncertainties before the fit to the data,
which are dominated by systematic uncertainties.
The $\chi^2/\mathrm{ndf}$ and the $\chi^2$ probability are also shown. Statistical uncertainties on MC predictions and data are uncorrelated across bins, while systematic uncertainties on the predictions are correlated.
}
\label{fig:leadJetReweight}
\end{figure}

\section{Analysis strategy}
To enhance the separation between signal and background,
a neural network algorithm (NN) is used.
Its architecture is sequential with two fully connected layers of 64 nodes,
and is implemented with the Python deep learning library,
Keras~\cite{chollet2015keras}.
The activation function used is the commonly employed `rectified linear unit' and the loss function is the `binary cross-entropy'.
Batch normalisation~\cite{ioffe2015batch} is performed to speed up the learning process,
dropout~\cite{hinton2012improving} is applied at a 10\% rate,
and the Adam algorithm~\cite{kingma2014adam} is used to optimise the parameters.

All signal samples are used in the training
against all background samples,
which are weighted according to their cross-sections.
The training is performed separately in each analysis region,
but the separate trainings include all \hplus\ mass samples,
and use the value of the \hplus\ mass as a parameter~\cite{Baldi:2016fzo}.
For signal events the parameter corresponds to the mass of the \hplus\ sample they belong to,
while for background events a random value of the \hplus\ mass,
taken from the distribution of signal masses,
is assigned to each event.
A total of 15 variables
are used in the NN:
\begin{itemize}
 
\item kinematic discriminant $D$ defined in the text,
\item scalar sum of the \pT\ of all jets,
\item centrality calculated using all jets and leptons,
\item \pT\ of the leading jet,
\item \pT\ of fifth leading jet,
\item invariant mass of the $b$-jet pair with minimum $\Delta R$,
\item invariant mass of the $b$-jet pair with maximum \pT,
\item largest invariant mass of a $b$-jet pair,
\item invariant mass of the jet triplet with maximum \pT,
\item invariant mass of the untagged jet-pair with minimum $\Delta R$ (not in 5j$\ge$4b),
\item average $\Delta R$ between all $b$-jet pairs in the event,
\item $\Delta R$ between the lepton and the pair of $b$-jets with smallest $\Delta R$,
\item second Fox--Wolfram moment calculated using all jets and leptons~\cite{Fox:1978vw},
\item number of jets (only in $\ge$6j3b and $\ge$6j$\ge$4b regions), and
\item number of $b$-jets (only in 5j$\ge$4b and $\ge$6j$\ge$4b regions).
 
\end{itemize}
The variables are chosen to provide the best discrimination against
the \ttbin\ background.
Among them,
the kinematic discriminant,
scalar sum of the \pT\ of all jets, centrality, and leading jet \pT\
provide the most discrimination.
The centrality is computed as the scalar sum of the \pT\ of all jets and leptons in the event
divided by the sum of their energies.
The kinematic discriminant is a variable reflecting the probability that an
event is compatible with the \hplustb\ hypothesis rather than the \ttbar\ hypothesis,
and is defined as
$D = P_{H^+(\textbf{x})} / ( P_{H^+(\textbf{x})} + P_{t\bar t(\textbf{x})} )$,
where $P_{H^+(\textbf{x})}$ and  $P_{t\bar t(\textbf{x})}$
are probability density functions for
$\textbf{x}$ under the signal hypothesis and background (\ttbar) hypothesis, respectively.
The event variable $\textbf{x}$ indicates the set of the missing transverse momentum and the four-momenta of the reconstructed lepton and the jets~\cite{HIGG-2017-04}.

Figure~\ref{fig:NNdiscrimination} shows the predicted NN output distributions in the four analysis regions
for selected \hplus\ signal samples and the SM background.
These distributions are used in a fit to extract the amount of \hplus\ signal in data.
The separation of the \hplus\ signal from the background is most difficult for low \hplus\ masses
because the two processes have very similar kinematics and topology.
The kinematic discriminant has large separating power at low \hplus\ masses,
whereas at higher masses,
where the topologies of the \hplus\ signal and the background are no longer alike,
other variables,
such as the scalar sum of the \pT\ of all jets,
provide the largest separation.
 
\begin{figure}[htb]
\centering
\subfloat[5j3b]{\includegraphics[width = 0.25\textwidth]{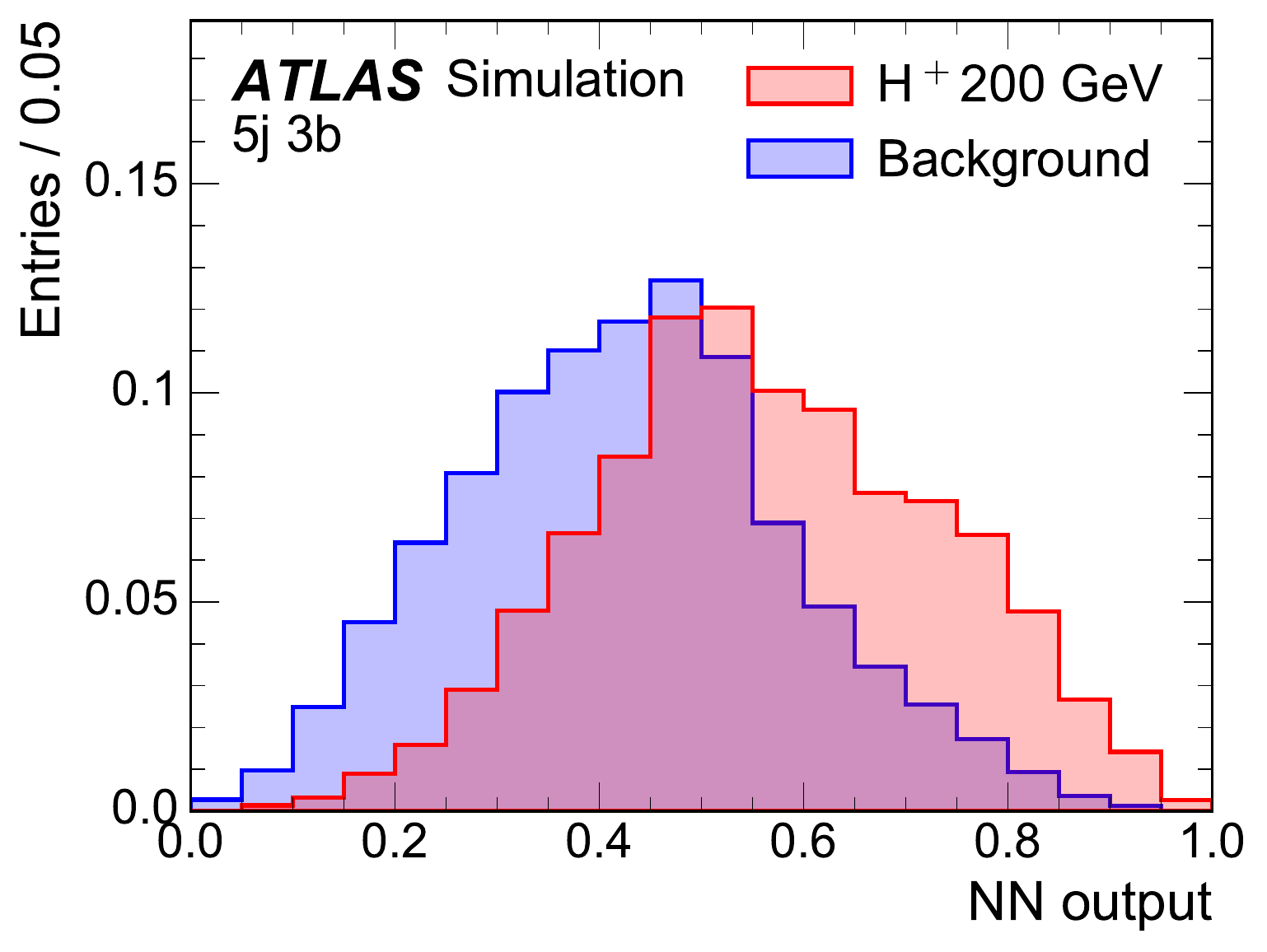}}
\subfloat[5j$\ge$4b]{\includegraphics[width = 0.25\textwidth]{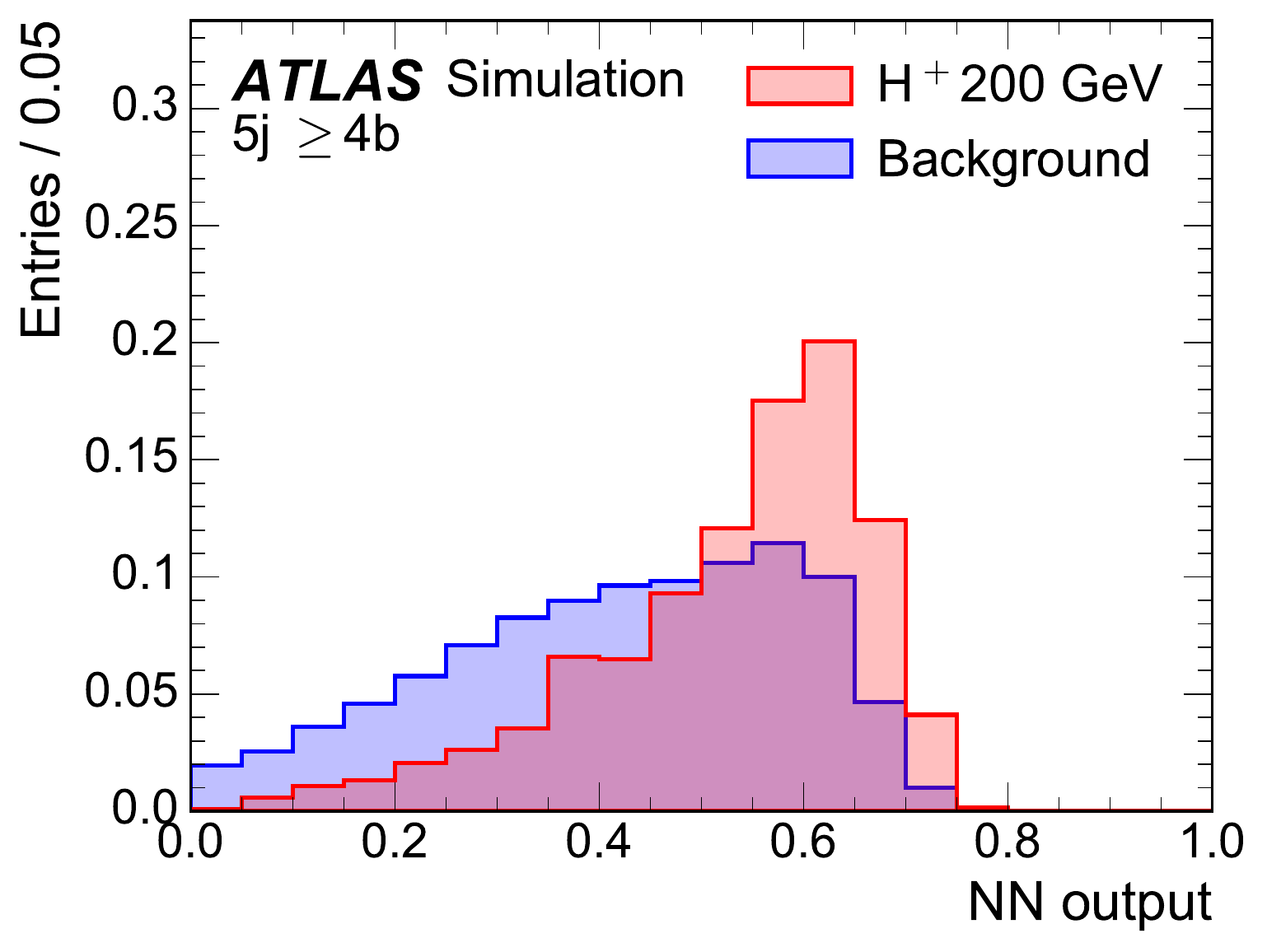}}
\subfloat[$\ge$6j3b]{\includegraphics[width = 0.25\textwidth]{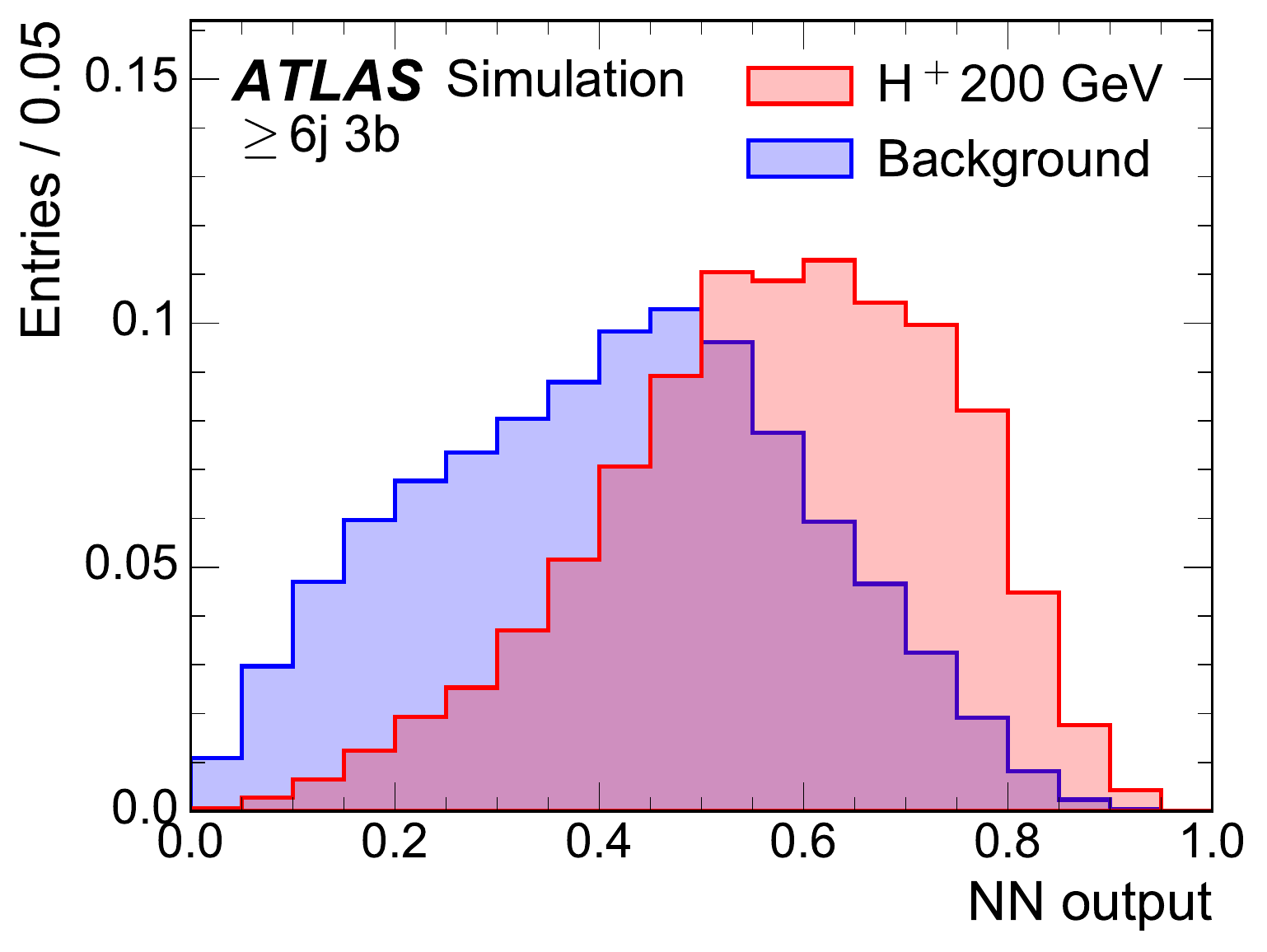}}
\subfloat[$\ge$6j$\ge$4b]{\includegraphics[width = 0.25\textwidth]{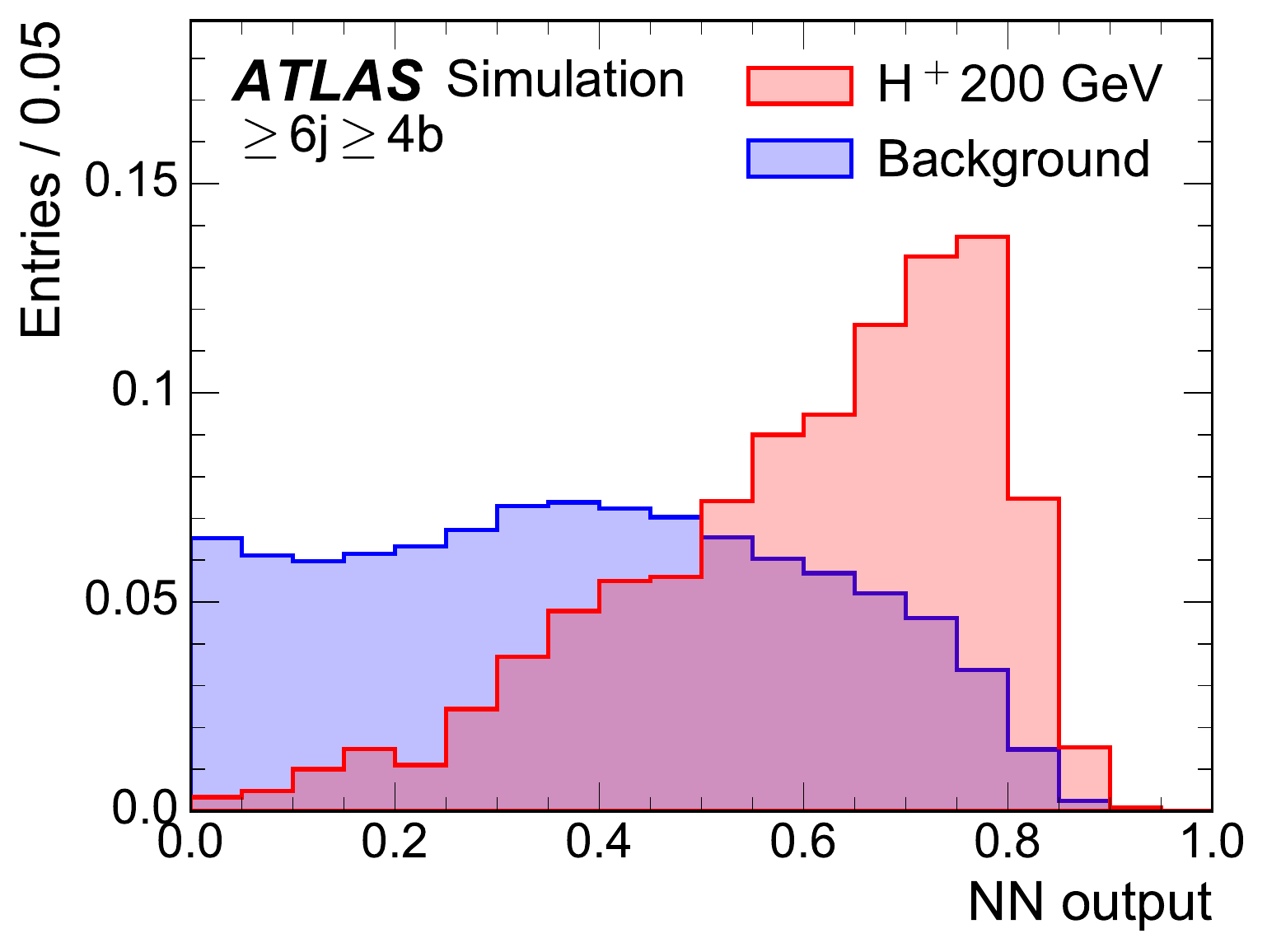}} \\
\subfloat[5j3b]{\includegraphics[width = 0.25\textwidth]{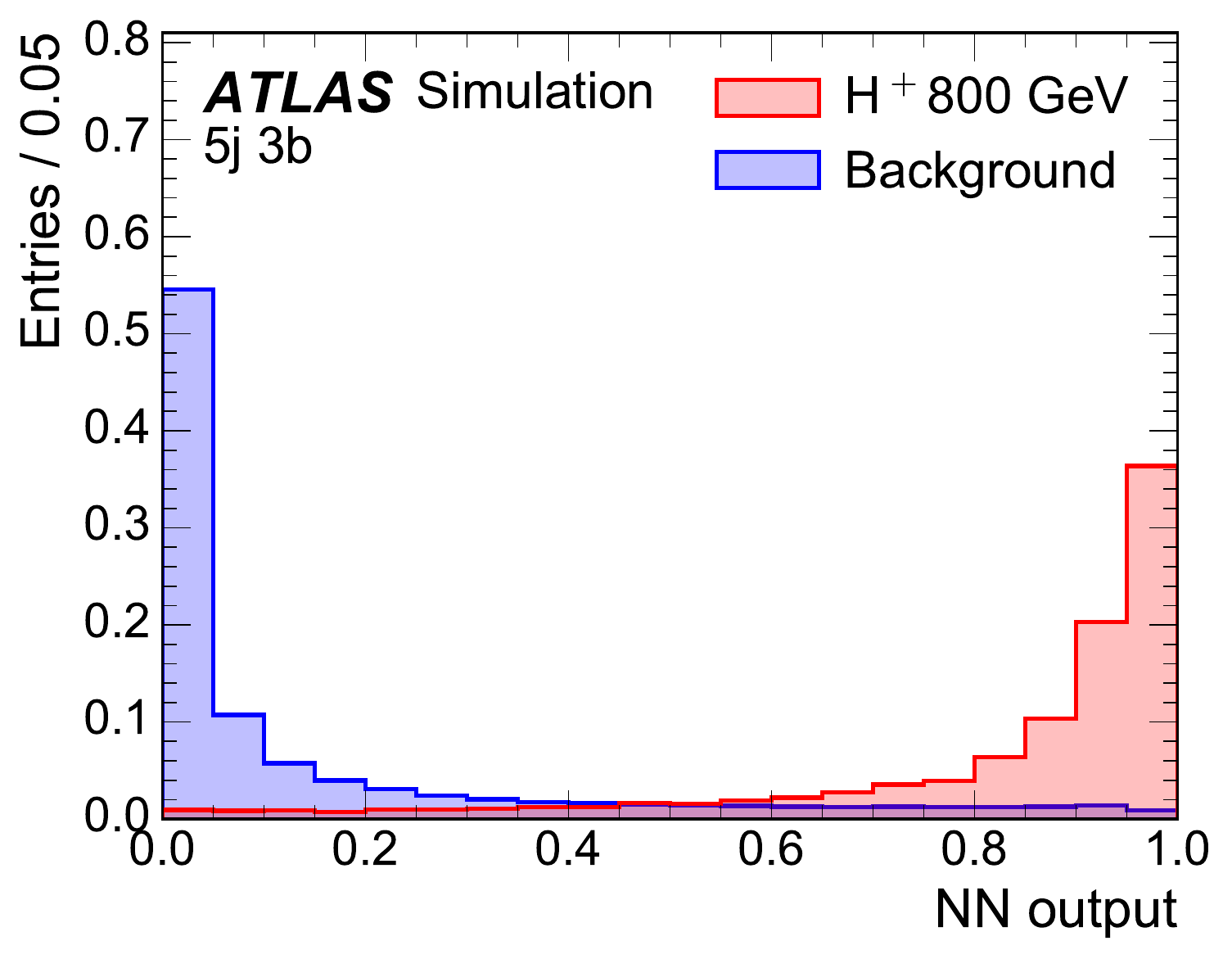}}
\subfloat[5j$\ge$4b]{\includegraphics[width = 0.25\textwidth]{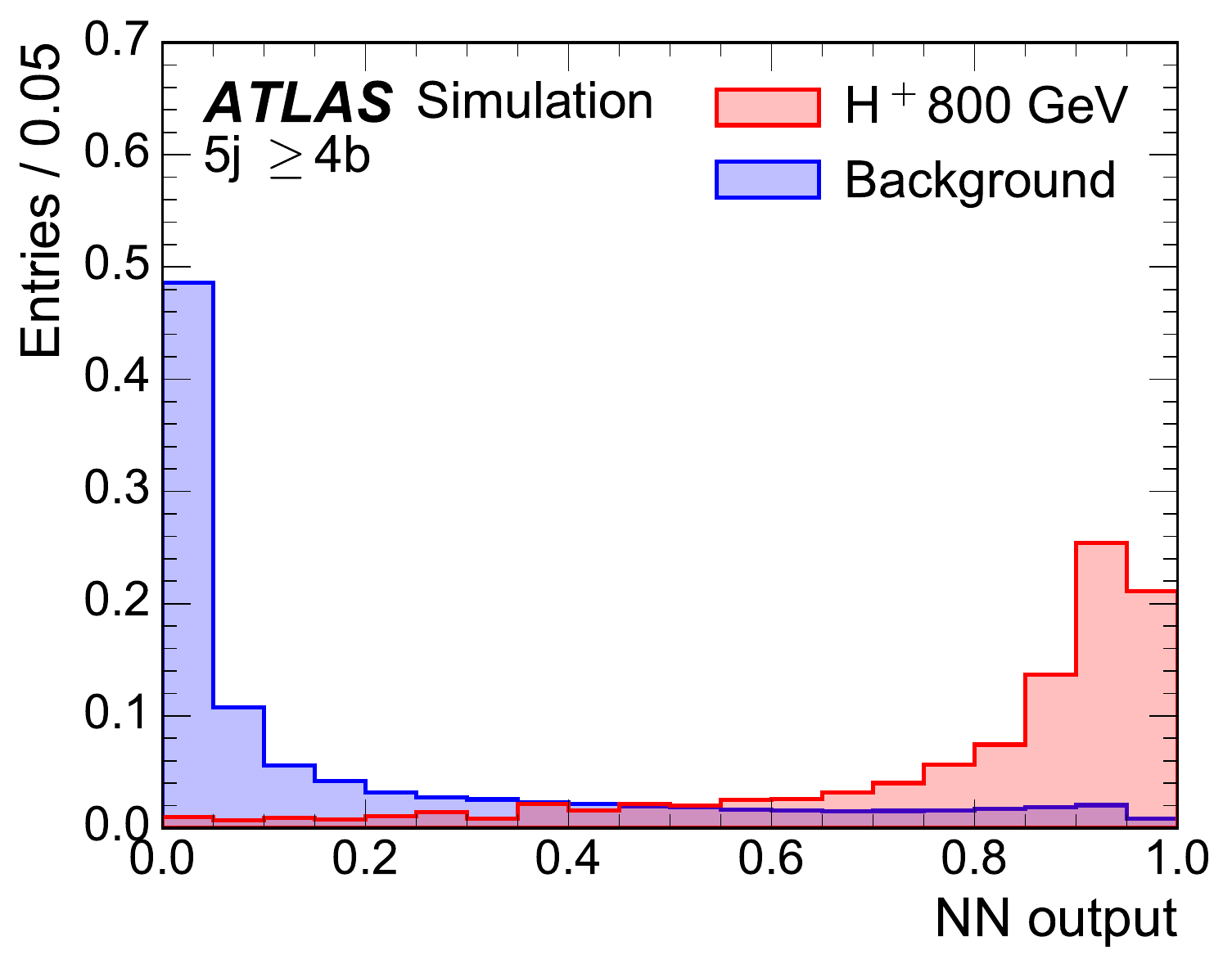}}
\subfloat[$\ge$6j3b]{\includegraphics[width = 0.25\textwidth]{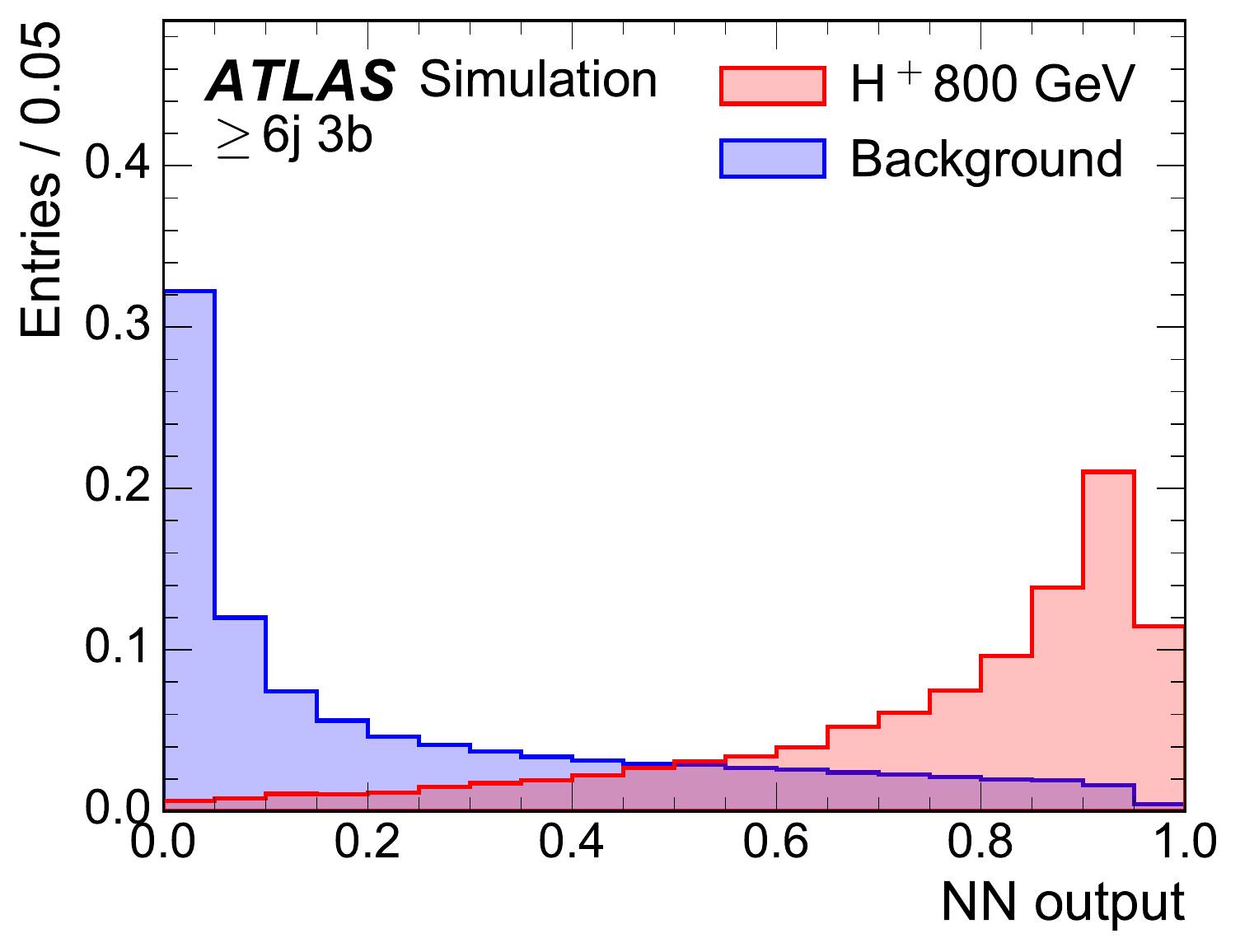}}
\subfloat[$\ge$6j$\ge$4b]{\includegraphics[width = 0.25\textwidth]{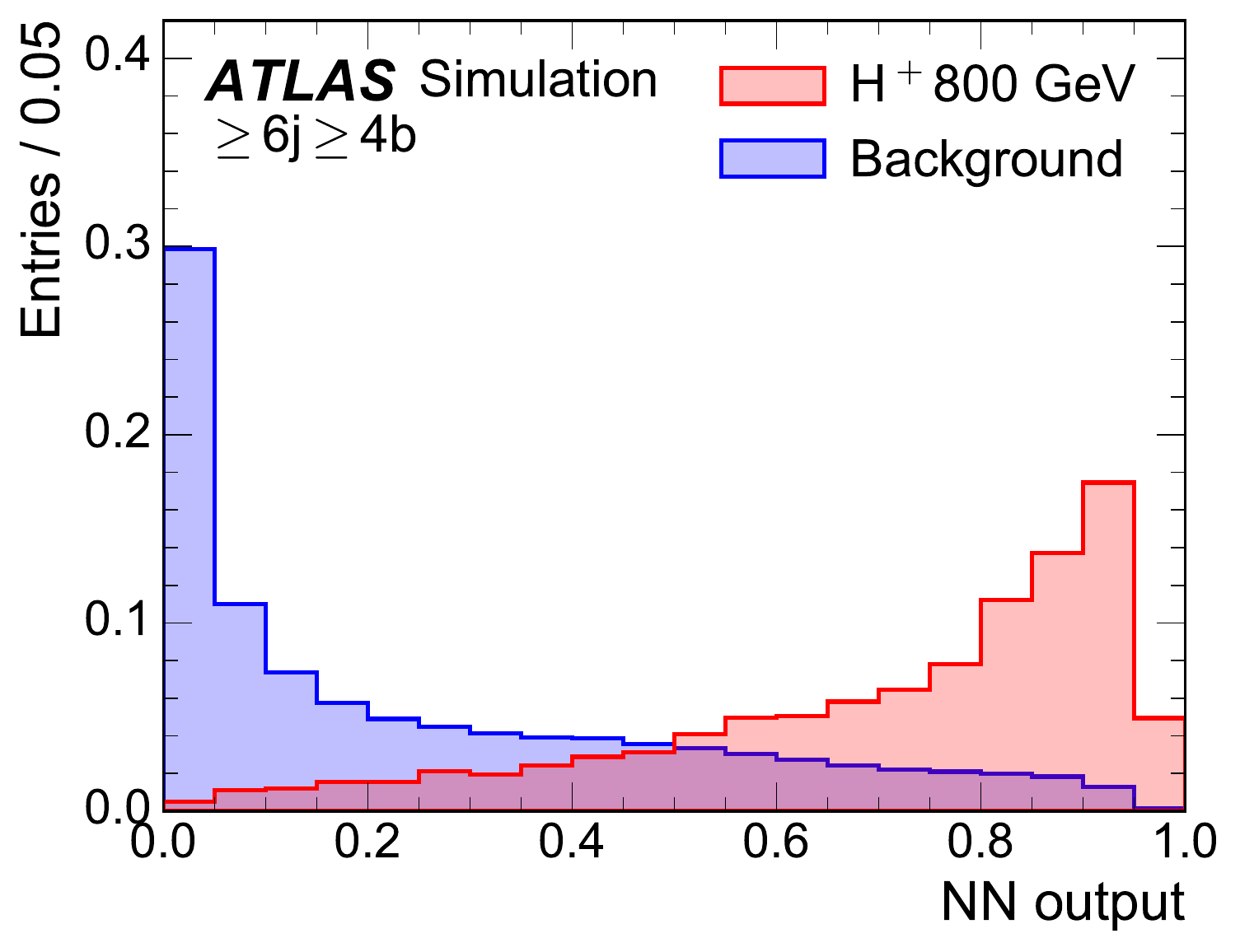}} \\
\caption{Expected distributions of the NN output for \hplus\ masses of 200~\GeV\ (top)
and 800~\GeV\ (bottom) for SM backgrounds and \hplus\ signal in the four analysis regions.
All distributions are normalised to unity.
}
\label{fig:NNdiscrimination}
\end{figure}

\section{Systematic uncertainties}
Various sources of experimental and theoretical uncertainties are considered in this analysis.
They may affect the overall normalisation of the processes,
the shapes of the NN output distributions, or both.
All the experimental uncertainties considered,
with the exception of that in the luminosity,
affect both normalisation and shape in all the simulated samples.
Uncertainties related to the modelling of the signal and background affect both normalisation and shape,
with the exception of cross-section uncertainties,
which only affect the normalisation of the sample considered.
Nonetheless, the normalisation uncertainties modify the relative fractions of the different samples,
leading to a shape variation in the final NN output distributions.
A single independent nuisance parameter (NP) is assigned to each source of
systematic uncertainty in the statistical analysis.
Some of the systematic uncertainties,
in particular most of the experimental uncertainties,
are decomposed into several independent sources.
Each individual source has a correlated effect across all
analysis regions and signal and background samples.
 
The uncertainty of the integrated luminosity for the full Run-2 dataset is 1.7\%~\cite{ATLAS-CONF-2019-021},
obtained using the LUCID-2 detector~\cite{LUCID2} for the primary luminosity measurements.
A variation in the pile-up reweighting of the simulated events described in Section~\ref{sec:dataMC}
is included to cover the uncertainty in the ratio of the predicted and measured inelastic cross-sections
in a given fiducial volume~\cite{STDM-2015-05}.
 
Uncertainties associated with charged leptons arise from the trigger selection,
the lepton reconstruction, identification and isolation criteria,
as well as the lepton momentum scale and resolution.
The reconstruction, identification and isolation efficiency of electrons and muons,
as well as the efficiency of the trigger used to record the events,
differ slightly between data and simulation,
which is compensated for by dedicated correction factors (CFs).
Efficiency CFs are measured using tag-and-probe techniques in $Z\rightarrow \ell^+ \ell^-$ data and simulated samples~\cite{PERF-2017-01,PERF-2015-10},
and are applied to the simulation to correct for the differences.
The effect of these CFs, as well as of their uncertainties,
are propagated as corrections to the MC event weight.
Additional sources of uncertainty originate from the corrections applied to adjust
the lepton momentum scale and resolution in the simulation to match those in data.
The impact of these uncertainties on $\sigma(pp \rightarrow \tbhplus) \times \BR(\hplustb)$ is smaller than 1\%.
 
Uncertainties associated with jets arise from the efficiency of pile-up rejection by the JVT,
from the jet energy scale (JES) and resolution (JER),
and from $b$-tagging.
Correction factors are applied to correct for differences between data and MC simulation for JVT efficiencies.
These CFs are estimated using $Z(\rightarrow \mu^+ \mu^-)$~+~jets
with tag-and-probe techniques similar to those in Ref.~\cite{PERF-2014-03}.
The JES and its uncertainty are derived by combining information from test-beam data,
collision data and simulation~\cite{PERF-2016-04}.
Additional uncertainties are considered,
related to jet flavour,
quark/gluon fraction,
pile-up corrections, $\eta$ dependence, high-\pT\ jets,
and differences between full and fast simulation.
The JER uncertainty is obtained by combining dijet balance measurements in data and simulation~\cite{JETM-2018-05}.
The $b$-tagging efficiencies in simulated samples are corrected to match efficiencies in data.
Correction factors are derived as a function of \pT\ for $b$-, $c$- and light-jets separately in dedicated calibration analyses.
For $b$-jet efficiencies, \ttbar\ events in the di-lepton topology are used,
exploiting the very pure sample of $b$-jets arising from the decays of the top quarks~\cite{FTAG-2018-01}.
For $c$-jet mistag rates,
\ttbar\ events in single-lepton topology are used,
exploiting the $c$-jets from the hadronically decaying $W$ bosons,
using techniques similar to those in Ref.~\cite{ATLAS-CONF-2018-001}.
For light-jet mistag rates,
the so-called negative-tag method similar to that in Ref.~\cite{ATLAS-CONF-2018-006} is used,
but using $Z$+jets events instead of di-jet events.
 
All the uncertainties described above on energy scales or resolutions of the reconstructed objects
are propagated to the missing transverse momentum.
Additional uncertainties in the scale and resolution of the soft term are considered,
which account for the disagreement between data and simulation of the \pT\ balance between
the hard and the soft components.
A total of three independent sources are added:
an offset along the hard component \pT\ axis,
and the smearing resolution along and perpendicular to this axis~\cite{PERF-2016-07,ATLAS-CONF-2018-023}.
Since the missing transverse momentum is not used in selection but only in the event reconstruction,
the associated uncertainties have an impact smaller than 1\%.
 
The uncertainty in the \hplus\ signal due to different scale choices is estimated by varying
\muf\ and \mur\ up and down by a factor of two.
The uncertainties from the modelling of the PDF are evaluated
replacing the nominal NNPDF2.3NLO PDF set by a symmetrised Hessian set,
PDF4LHC15\_nlo\_30,
following the PDF4LHC recommendations for LHC Run 2~\cite{Butterworth:2015oua}.
 
The modelling of the \ttbar+~jets background is one of the largest sources of uncertainty in the analysis,
and several different components are considered.
The 6\% uncertainty for the inclusive \ttbar\ production cross-section predicted at NNLO+NNLL
includes effects from varying \muf\ and \mur,
the PDF, $\alpha_{\textrm{S}}$, and the top-quark mass~\cite{LHCTopWGTTbar}.
This uncertainty is applied to \ttbar~+~light only,
since the normalisation of \ttbin\ and \ttcin\ are allowed to vary freely in the fit.
Besides normalisation,
the \ttlin, \ttbin\ and \ttcin\ processes are affected by different types of uncertainties:
the uncertainties associated with additional Feynman diagrams for the \ttlin\ are constrained from relatively
precise measurements in data~\cite{TOPQ-2017-01};
\ttbin\ and \ttcin\ can have similar or different Feynman diagrams
depending on the flavour scheme used for the PDF,
and the different masses of the $b$- and the $c$-quarks contribute to additional differences between these two processes.
For these reasons,
all uncertainties in the \ttbar~+~jets background modelling
are assigned independent NP for \ttlin, \ttbin\ and \ttcin.
Systematic uncertainties in the acceptance and shapes are extracted by comparing
the nominal prediction with alternative MC samples or settings.
Such comparisons would significantly change the fractions of \ttbin\ and \ttcin.
However,
since the normalisation of these sub-processes in the analysis regions
is determined in the fit,
these alternative predictions are reweighted in such a way
that they keep the same fractions of \ttbin\ and \ttcin\ as the nominal sample
in the phase-space selected by the analysis.
 
The uncertainty due to initial state radiation (ISR) is estimated by using the \textit{Var3cUp} (\textit{Var3cDown}) variant from the A14 tune~\cite{ATL-PHYS-PUB-2014-021},
corresponding to $\alpha_{\textrm{S}}^{\mathrm{ISR}}=$~0.140 (0.115) instead of the nominal $\alpha_{\textrm{S}}^{\mathrm{ISR}}=0.127$.
Uncertainties related to \muf\ and \mur\ are estimated by scaling each one up and down by a factor of two.
For the final state radiation (FSR),
the amount of radiation is increased (decreased) in the PS
corresponding to $\alpha_{\textrm{S}}^{\mathrm{FSR}}=$~0.142 (0.115) instead of the nominal
$\alpha_{\textrm{S}}^{\mathrm{FSR}}=0.127$.
The nominal \powheg+\pythia\ sample is compared with the \powheg+\herwig\ sample to assess the effect
of the PS and hadronisation models,
and to the \madgraph\ sample to assess the effect of the NLO matching technique.
The nominal \powheg+\pythia\ 5FS prediction for the dominant \ttbin\ background,
in which all the additional partons are produced by the PS,
is compared with an alternative \powheg+\pythia\ 4FS sample,
in which the \bbbar\ pair is generated in addition to the \ttbar\ pair at the ME level.
An uncertainty resulting from the comparison of the shapes of the two models is included.
Finally,
the weights derived in Section~\ref{sec:background} to improve the agreement of the simulation with data
are varied within their statistical uncertainties, in a correlated way among the three \ttbar~+~jets components.
All the sources of systematic uncertainty for the \ttbar~+~jets modelling are summarised in Table~\ref{tab:ttjetsSysts}.
 
\begin{table}[htbp]
\caption{
Summary of the sources of systematic uncertainty for \ttbar~+~jets modelling.
The systematic uncertainties listed in the second section of the table are evaluated in such a way as
to have no impact on the normalisation of the three,
\ttbin, \ttcin\ and \ttlin,
components in the phase-space selected in the analysis.
The last column of the table indicates the \ttbar~+~jets components to which the systematic uncertainty is assigned.
All systematic uncertainty sources,
except those associated to the \ttbar\ reweighting,
are treated as uncorrelated across the three components.
}
\begin{center}
\small
\begin{tabular}{ llll }
\hline
Uncertainty source      & \multicolumn{2}{l}{Description}                                 &   Components    \\
\hline\hline
\ttbar\ cross-section   & \multicolumn{2}{l}{Up or down by 6\%}                           &   \ttlin\       \\
\ttbar\ reweighting     & \multicolumn{2}{l}{Statistical uncertainties of fitted function (six) parameters}        &   All \ttbar\ and $Wt$ \\
\ttbin\ modelling       & \multicolumn{2}{l}{4FS vs 5FS}                                  &   \ttbin\       \\
\ttbin\ normalisation   & \multicolumn{2}{l}{Free-floating}                               &   \ttbin\       \\
\ttcin\ normalisation   & \multicolumn{2}{l}{Free-floating}                               &   \ttcin\       \\
\hline
NLO matching            & \madgraph+\pythia                      & vs  \powheg+\pythia    &   All \ttbar\   \\
PS \& hadronisation     & \powheg+\herwig                        & vs  \powheg+\pythia    &   All \ttbar\   \\
ISR            & Varying $\alpha_{\textrm{S}}^{\mathrm{ISR}}$    & in  \powheg+\pythia    &   All \ttbar\   \\
\muf\                   & Scaling by 0.5 (2.0)                   & in  \powheg+\pythia    &   All \ttbar\   \\
\mur\                   & Scaling by 0.5 (2.0)                   & in  \powheg+\pythia    &   All \ttbar\   \\
FSR            & Varying $\alpha_{\textrm{S}}^{\mathrm{FSR}}$    & in  \powheg+\pythia    &   All \ttbar\   \\
\hline
\end{tabular}
\end{center}
\label{tab:ttjetsSysts}
\end{table}
 
A 5\% uncertainty is considered for the cross-sections of the three single-top production modes~\cite{LHCTopWGSGtop,Martin:2009iq,Martin:2009bu,Aliev:2010zk,Kant:2014oha}.
Uncertainties associated with the PS and hadronisation model,
and with the NLO matching scheme are evaluated by comparing, for each process,
the nominal \powheg+\pythia\ sample with a sample produced using \powheg+\herwig\ \cite{Bellm:2015jjp} and \madgraph+\pythia, respectively.
As mentioned in Section~\ref{sec:background},
the $Wt$ single-top mode is included in the reweighting procedure,
and thus the same uncertainties used for \ttbar\ are applied here.
The uncertainty associated to the interference between $Wt$ and \ttbar\ production at NLO~\cite{Frixione:2008yi}
is assessed by comparing the nominal \powheg+\pythia\ sample produced using the ``diagram removal''
scheme with an alternative sample produced with the same generator
but using the ``diagram subtraction'' scheme~\cite{Frederix:2012dh,Frixione:2008yi}.
 
The predicted SM $\ttbar H$ signal cross-section uncertainty is
$^{+5.8\%}_{-9.2\%}$ (QCD scale) $\pm3.6\%$ (PDF + $\alpha_{\mathrm{S}}$)~\cite{deFlorian:2016spz,Raitio:1978pt,Beenakker:2002nc,Dawson:2003zu,Yu:2014cka,Frixione:2015zaa}.
Uncertainties of the Higgs boson branching ratios
amount to 2.2\% for the $b\bar{b}$ decay mode~\cite{deFlorian:2016spz}.
For the ISR and FSR, the amount of radiation is varied following the same procedure as for \ttbar.
The nominal \powheg+\pythia\ sample is compared with the \powheg+\herwig\ sample to assess the uncertainty due to
PS and hadronisation,
and to the \madgraph\ sample for the uncertainty due to the NLO matching.
 
The uncertainty of the $\ttbar V$ NLO cross-section prediction is 15\%,
split into PDF and scale uncertainties as for
$\ttbar H$~\cite{deFlorian:2016spz,Campbell:2012dh}.
An additional $\ttbar V$ modelling uncertainty,
related to the choice of PS and hadronisation model and matching scheme,
is assessed by comparing the nominal \madgraph+\pythia\ samples with alternative samples generated with \sherpa.
 
An overall 50\% normalisation uncertainty is considered for the four-top-quarks background,
covering effects from varying \muf, \mur, PDF and $\alpha_{\textrm{S}}$~\cite{Frederix:2017wme,Alwall:2014hca}.
The small background $tZq$ is assigned
a 7.9\% and a 0.9\% uncertainty accounting for \muf\ and \mur\ scales
and PDF variations, respectively.
Finally, a single 50\% uncertainty is used for $tZW$~\cite{Alwall:2014hca}.
 
An uncertainty of 40\% is assumed for the $W$+jets normalisation,
with an additional 30\% for $W$ + heavy-flavour jets,
taken as uncorrelated between events with two and more than two heavy-flavour jets.
These uncertainties are based on variations of the \muf\ and \mur\ scales
and of the matching parameters in the \sherpa\ samples.
An uncertainty of 35\% is applied to the $Z$+jets normalisation,
uncorrelated across jet bins,
to account for both the variations of the scales and matching parameters in the
\sherpa\ samples and the uncertainty in the extraction from data of the correction factor for the heavy-flavour component~\cite{Bothmann_2016,Ball:2014uwa}.
Finally, a 50\% normalisation uncertainty in the diboson background is assumed,
which includes uncertainties in the inclusive cross-section and additional jet production~\cite{ATL-PHYS-PUB-2016-002}.
 
\section{Results}
\label{sec:results}
A binned maximum-likelihood fit to the data is performed simultaneously
on the NN output distributions in the four analysis regions,
and each mass hypothesis is tested separately.
A profile-likelihood-ratio test and the CL$_{\textrm S}$ method~\cite{Cowan:2010js,Read:2002hq,Junk_1999} are
used to quantify the level of agreement with the background-only hypothesis or background-plus-signal hypothesis
and to determine exclusion limits.
The parameter of interest is the product of
the production cross-section $\sigma (pp \rightarrow \tbhplus)$
and the branching ratio \BR(\hplustb).
In addition,
two unconstrained factors, for which no prior knowledge is assumed, are used to normalise the \ttbin\ and \ttcin\ backgrounds.
These normalisation factors range from 1.2 to 1.6 (0.2 to 1.8)
with a typical uncertainty of 0.2 (0.6) for the \ttbin\ (\ttcin) background,
depending on the \hplus\ mass hypothesis used in the fit. The fitted \ttbar\ + heavy-flavour jets backgrounds in the different fits are consistent within two standard deviations.
All systematic uncertainties are implemented as nuisance parameters with log-normal constraint terms.
There are about 170 nuisance parameters considered in the fit,
the number varying slightly across the range of mass hypotheses.
A summary of the systematic uncertainties with similar sources grouped together is
given in Table~\ref{tab:systematics}.
Depending on the particular \hplus\ mass hypothesis,
the total systematic uncertainty is dominated by the uncertainties in the modelling of the
\ttbar\ background,
in particular \ttbin\ and \ttcin,
jet flavour-tagging uncertainties,
and jet energy scale and resolution.
 
\begin{table}[htb]
\caption{
Summary of the statistical and systematic uncertainties on
$\mu = \sigma(pp \rightarrow \tbhplus) \times \BR(\hplustb)$ shown for a \hplus\
signal with a mass of 200 and 800~\GeV,
extracted from the fit to the data. A value of $\mu = 1$~pb is assumed for all \hplus\ mass hypotheses.
Due to correlations between the different sources of uncertainty,
the total systematic uncertainty can be different from the sum in quadrature of the individual sources.
The normalisation factors for both \ttbin\ and \ttcin\ are included in the statistical component.\vspace{0.5cm}}
\begin{center}
\begin{tabular}{l c c}
\hline
Uncertainty source   & $\Delta\mu(\hplus_{200})$ [pb] & $\Delta\mu(\hplus_{800})$ [pb] \\
\hline \hline
\ttbin\ modelling                            &  1.01   &  0.025   \\
Jet energy scale and resolution            &  0.35   &  0.009  \\
\ttcin\ modelling                          &  0.32   &  0.006   \\ 
Jet flavour tagging                        &  0.20   &  0.025   \\
Reweighting                                &  0.22   &  0.007   \\ 
\ttlin\ modelling                          &  0.33   &  0.009   \\ 
Other background modelling                 &  0.19   &  0.011   \\
MC statistics                              &  0.11   &  0.008   \\ 
JVT, pile-up modelling                     &  <0.01~~   &  0.001   \\ 
Luminosity                                 &  <0.01~~  &  0.002   \\ 
Lepton ID, isolation, trigger, \MET        &  <0.01~~  &  <0.001~~  \\ 
\hplus\ modelling                          &  0.05   &  0.002    \\ 
\hline
Total systematic uncertainty               &  1.35   &  0.049    \\ 
\hline
\ttbin\ normalisation                      &  0.23   &  0.007   \\ 
\ttcin\ normalisation                      &  ~~0.045  &  0.015   \\
\hline
Total statistical uncertainty              &  0.43   &  0.025   \\ 
\hline \hline
Total uncertainty                          &  1.42   &  0.055   \\ 
\hline
\end{tabular}
\end{center}
\label{tab:systematics}
\end{table}
 
Table~\ref{tab:postfitYields} shows the event yields after the background-plus-signal fit
under the 200~\GeV\ and 800~\GeV\ \hplus\ mass hypotheses.
The corresponding post-fit distributions of the NN output in each analysis region
are shown in Figure~\ref{fig:NNpostfit} for the 200~\GeV\ and 800~\GeV\ \hplus\ mass hypotheses.
After the fit,
good agreement between the data and simulation is found in the input variables to the NN.
No significant excess above the expected SM background is observed in all regions and mass intervals and
upper limits on the cross-section times branching ratio are derived as function of the \hplus\ mass.
 
\begin{table}[!htb]
\caption{
Event yields of the \hplus\ signal and SM background processes in the four analysis regions
after the fit to the data under the \hplus\ mass (\mhp)  hypotheses of 200~\GeV\ (top) and 800~\GeV\ (bottom).
The quoted uncertainties take into account correlations and constraints of the nuisance parameters
and include both the statistical and systematic uncertainties.
The signal yield uncertainty includes the uncertainty of the
$\sigma(pp \rightarrow \tbhplus) \times \BR(\hplustb)$ values
fitted under the 200 or 800~\GeV\ \hplus\ mass hypotheses.
Negative correlations among \ttlin, \ttbin\ and \ttcin\ modelling uncertainties
can cause the uncertainty on the total yields to be smaller than on individual components. \vspace{0.5cm}
}
\begin{center}
\begin{tabular}{l c c c c}
\hline
\multicolumn{5}{c}{\mhp = 200~\GeV\ hypothesis}  \\
\hline
\hline
& {5j, 3b} & {5j, $\geq$ 4b} & {$\geq$ 6j, 3b} & {$\geq$ 6j, $\geq$ 4b}   \\
\hline
\ttlin                  & 45000 $\pm$ 4000 & ~~310 $\pm$ 110 & 32000 $\pm$ 4000 & ~~340 $\pm$ 140  \\
\ttbin                  & 29600 $\pm$ 2900 & 2940 $\pm$ 220 & 40200 $\pm$ 3300 & 8000 $\pm$ 500  \\
\ttcin                  & 14000 $\pm$ 4000 & ~~440 $\pm$ 140 & 19000 $\pm$ 6000 & 1010 $\pm$ 290  \\
$t\bar{t}$ + $W$        & 110 $\pm$ 15 & ~~3.2 $\pm$ 0.6 & 236 $\pm$ 35 & 16.2 $\pm$ 2.7  \\
$t\bar{t}$ + $Z$        & 300 $\pm$ 40 & 51 $\pm$ 6 & 670 $\pm$ 90 & 174 $\pm$ 23  \\
Single-top $Wt$-channel & 2300 $\pm$ 600 & ~~80 $\pm$ 50 & 1900 $\pm$ 800 & 150 $\pm$ 90  \\
Single-top $t$-channel & ~~740 $\pm$ 300 & ~~51 $\pm$ 20 & ~~500 $\pm$ 400 & ~~60 $\pm$ 50  \\
Other top-quark sources       & 128 $\pm$ 16 & 17.5 $\pm$ 3.2 & 180 $\pm$ 70 & ~~58 $\pm$ 24 \\
$VV$ \& $V$ + jets      & 1600 $\pm$ 600 & ~~65 $\pm$ 23 & 1600 $\pm$ 600 & 120 $\pm$ 40  \\
$t\bar{t}H$             & 530 $\pm$ 60 & 127 $\pm$ 19 & 1140 $\pm$ 120 & 430 $\pm$ 60  \\
\hline
\hplus\                   & ~~600 $\pm$ 900 & ~~70 $\pm$ 90 & ~~~~700 $\pm$ 1000 & ~~160 $\pm$ 230  \\
\hline
Total                   & 95700 $\pm$ 2900 & 4150 $\pm$ 140 & 98400 $\pm$ 2900 & 10500 $\pm$ 400~~  \\
\hline
\hline
Data                    & 95852             & 4109             & 98929             & 10552    \\
\noalign{\vskip 1cm}
\hline
\multicolumn{5}{c}{\mhp = 800~\GeV\ hypothesis}   \\
\hline
\hline
& {5j, 3b} & {5j, $\geq$ 4b} & {$\geq$ 6j, 3b} & {$\geq$ 6j, $\geq$ 4b}   \\
\hline
\ttlin                  & 46000 $\pm$ 4000 & ~~330 $\pm$ 120 & 33000 $\pm$ 4000 & ~~500 $\pm$ 200   \\
\ttbin                  & 29600 $\pm$ 3100 & 2920 $\pm$ 210 & 41000 $\pm$ 4000 & 8100 $\pm$ 400  \\
\ttcin                  & 14000 $\pm$ 6000 & ~~440 $\pm$ 190 & 17000 $\pm$ 7000 & ~~870 $\pm$ 330   \\
$t\bar{t}$ + $W$        & 108 $\pm$ 15 & ~~3.3 $\pm$ 0.6 & 233 $\pm$ 35 & 16.0 $\pm$ 2.7  \\
$t\bar{t}$ + $Z$        & 300 $\pm$ 40 & 50 $\pm$ 7 & 660 $\pm$ 90 & 171 $\pm$ 23  \\
Single-top $Wt$-channel & 2000 $\pm$ 500 & ~~56 $\pm$ 33 & 1400 $\pm$ 500 & 100 $\pm$ 60   \\
Single-top $t$-channel  & ~~740 $\pm$ 300 & ~~53 $\pm$ 21 & ~~600 $\pm$ 500 & ~~70 $\pm$ 50  \\
Other top-quark sources       & 130 $\pm$ 16 & 17.7 $\pm$ 3.2 & 190 $\pm$ 70 & ~~61 $\pm$ 24 \\
$VV$ \& $V$ + jets      & 1900 $\pm$ 700 & ~~73 $\pm$ 25 & 1700 $\pm$ 600 & 130 $\pm$ 50   \\
$t\bar{t}H$             & 520 $\pm$ 60 & 125 $\pm$ 19 & 1130 $\pm$ 120 & 420 $\pm$ 60 \\
\hline
\hplus\                 & ~~30 $\pm$ 80 & ~~~~4 $\pm$ 10 & ~~~~70 $\pm$ 180 & ~~20 $\pm$ 50   \\
\hline
Total                   & 94700 $\pm$ 2800 & 4070 $\pm$ 140 & 97800 $\pm$ 2800 & 10400 $\pm$ 400~~ \\
\hline
\hline
Data                    & 95852             & 4109             & 98929             & 10552 \\
\hline
\end{tabular}
\end{center}
\label{tab:postfitYields}
\vspace{1cm}
\end{table}

\begin{figure}[htb]
\centering
\subfloat[5j3b]
{\includegraphics[width = 0.25\textwidth]{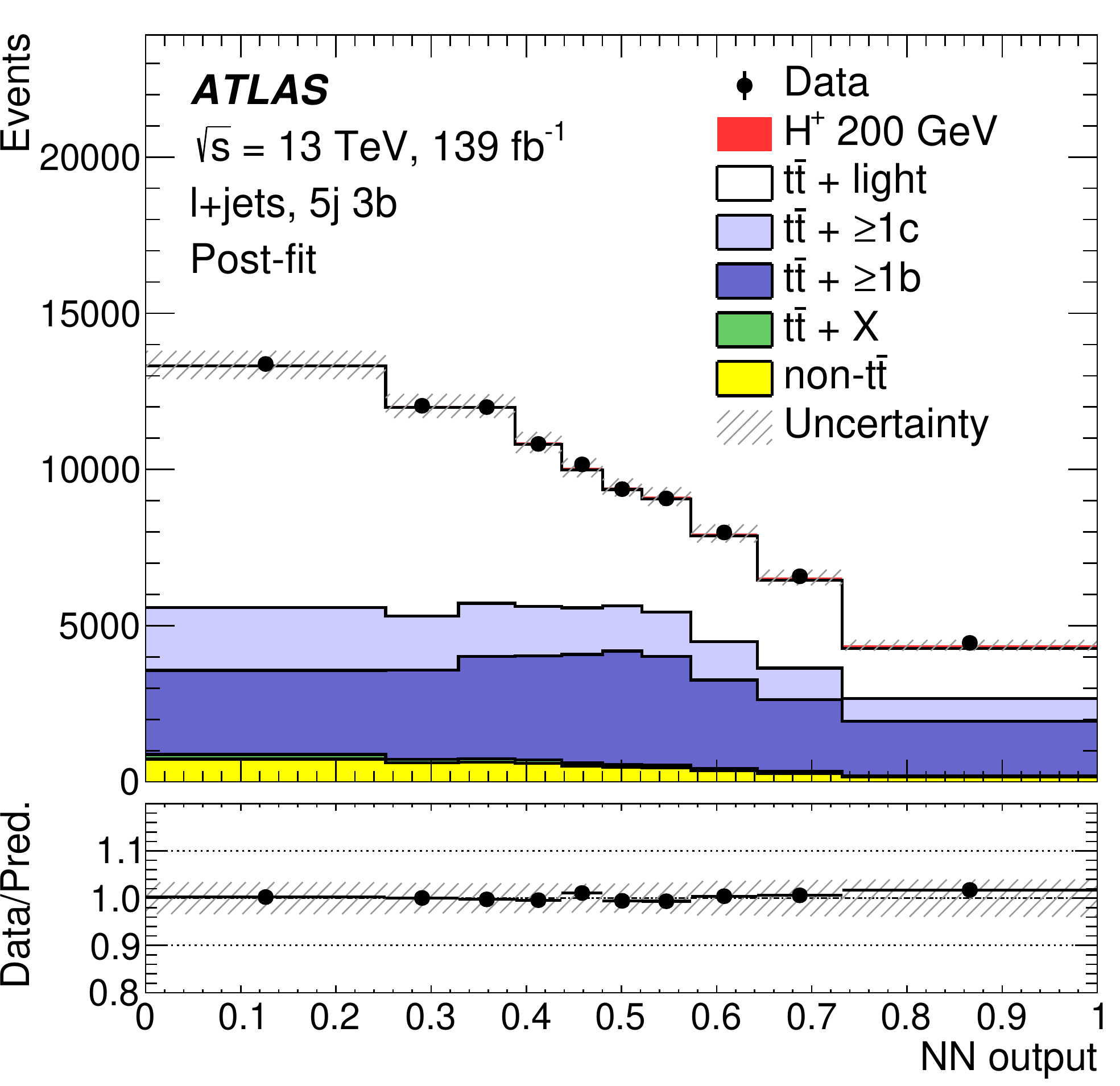}}
\subfloat[5j$\geq$4b]
{\includegraphics[width = 0.25\textwidth]{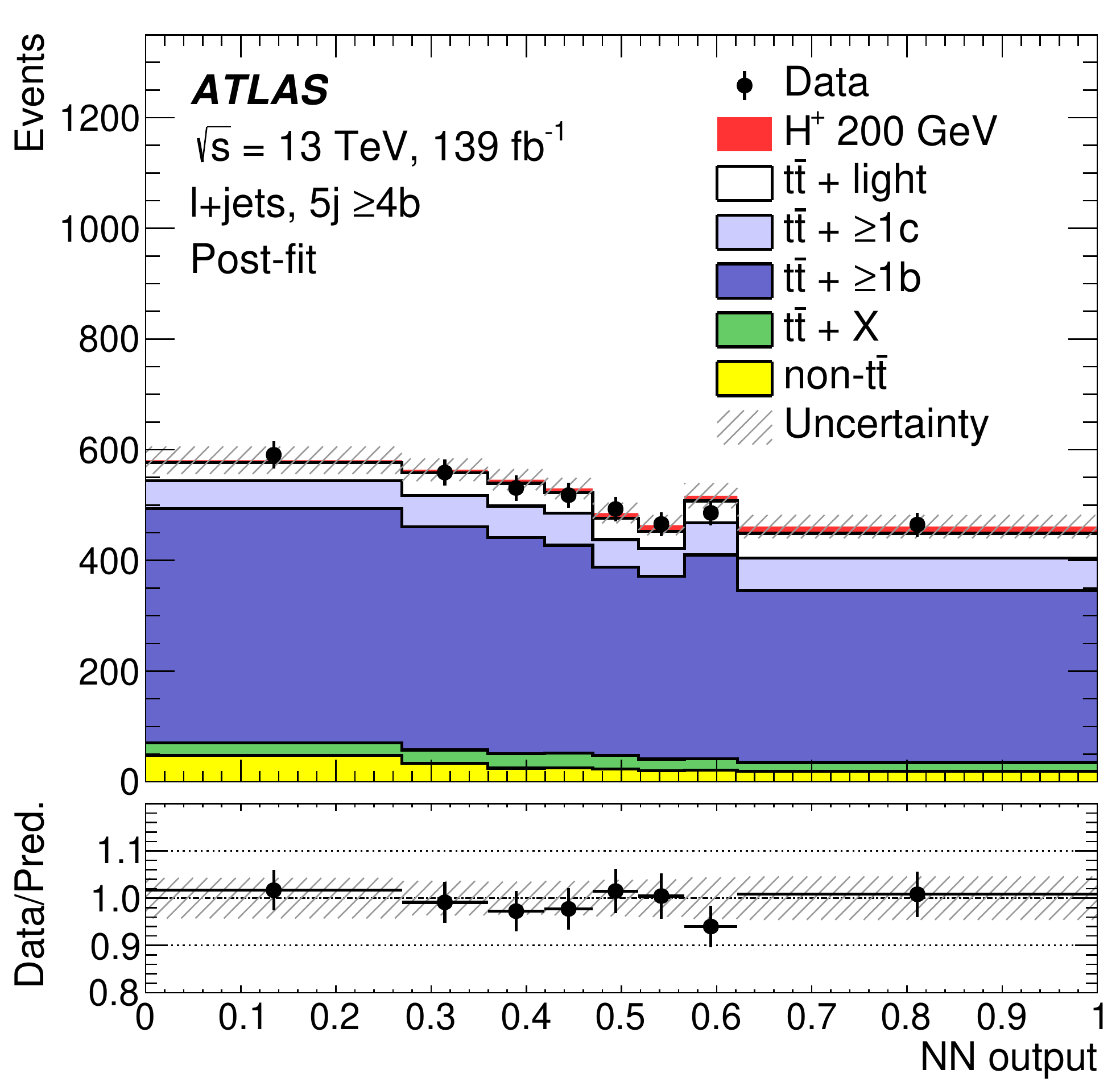}}
\subfloat[$\geq$ 6j3b]
{\includegraphics[width = 0.25\textwidth]{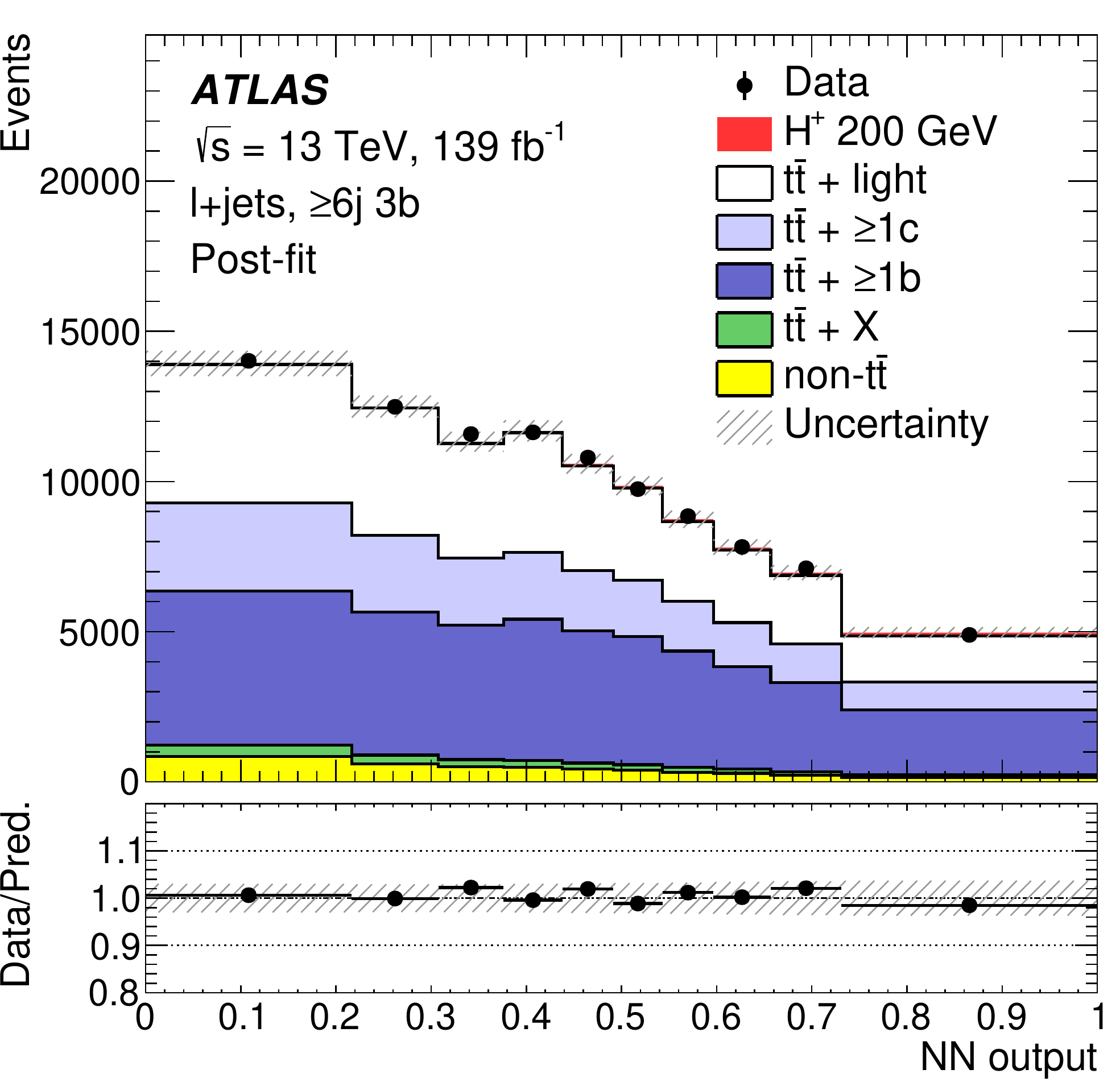}}
\subfloat[$\geq$6j$\geq$4b]
{\includegraphics[width = 0.25\textwidth]{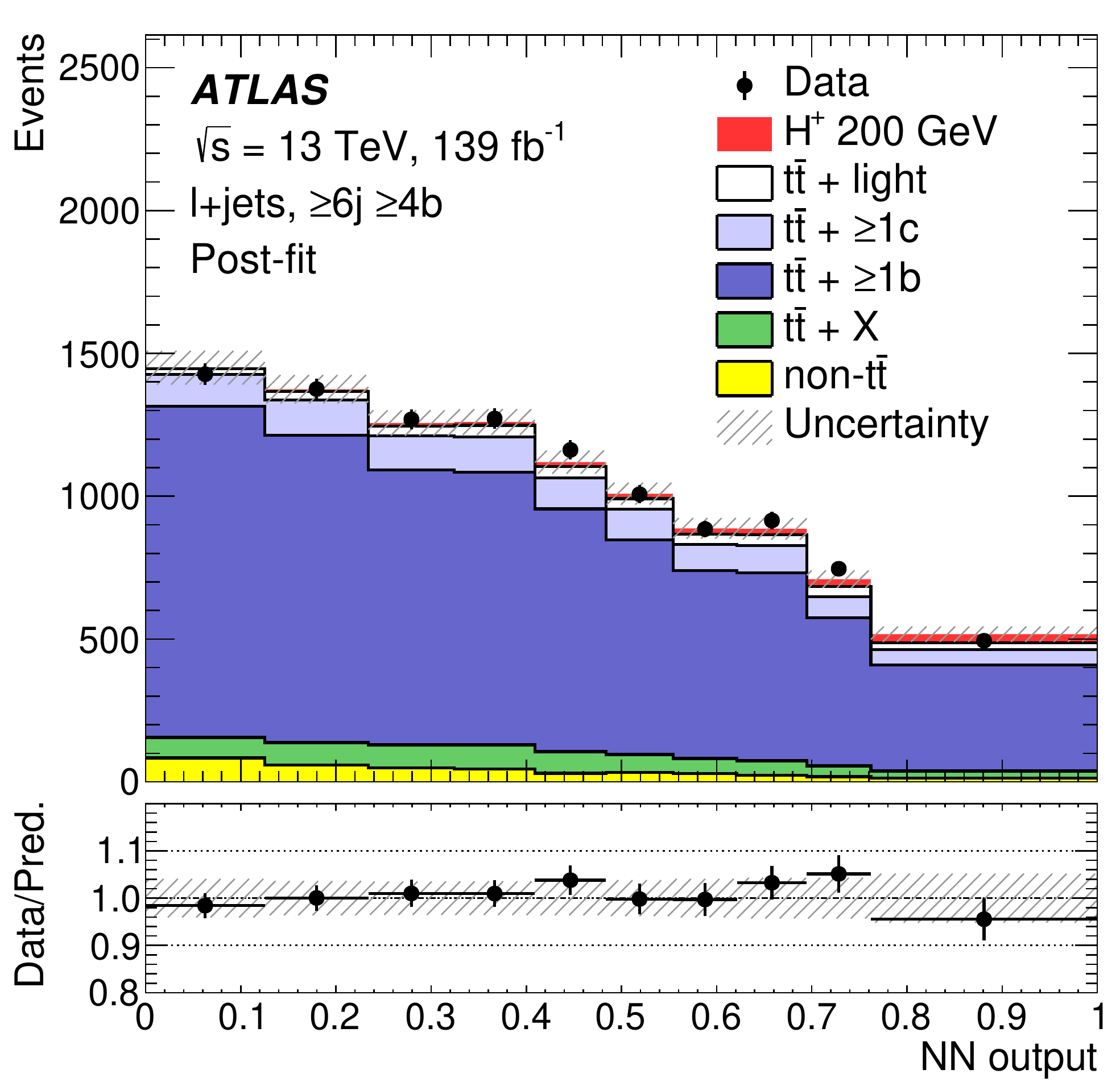}} \\
\subfloat[5j3b]
{\includegraphics[width = 0.25\textwidth]{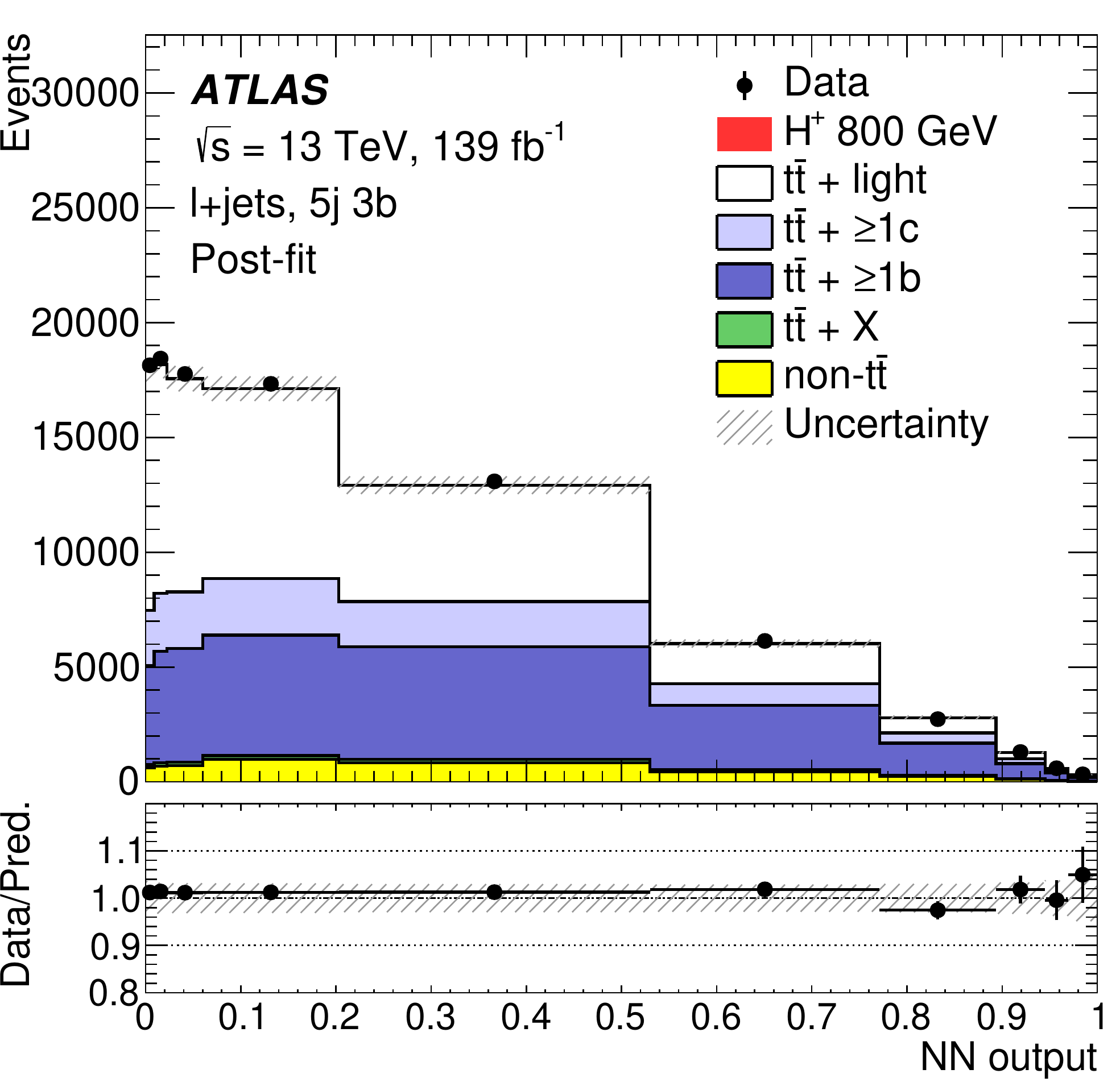}}
\subfloat[5j$\geq$4b]
{\includegraphics[width = 0.25\textwidth]{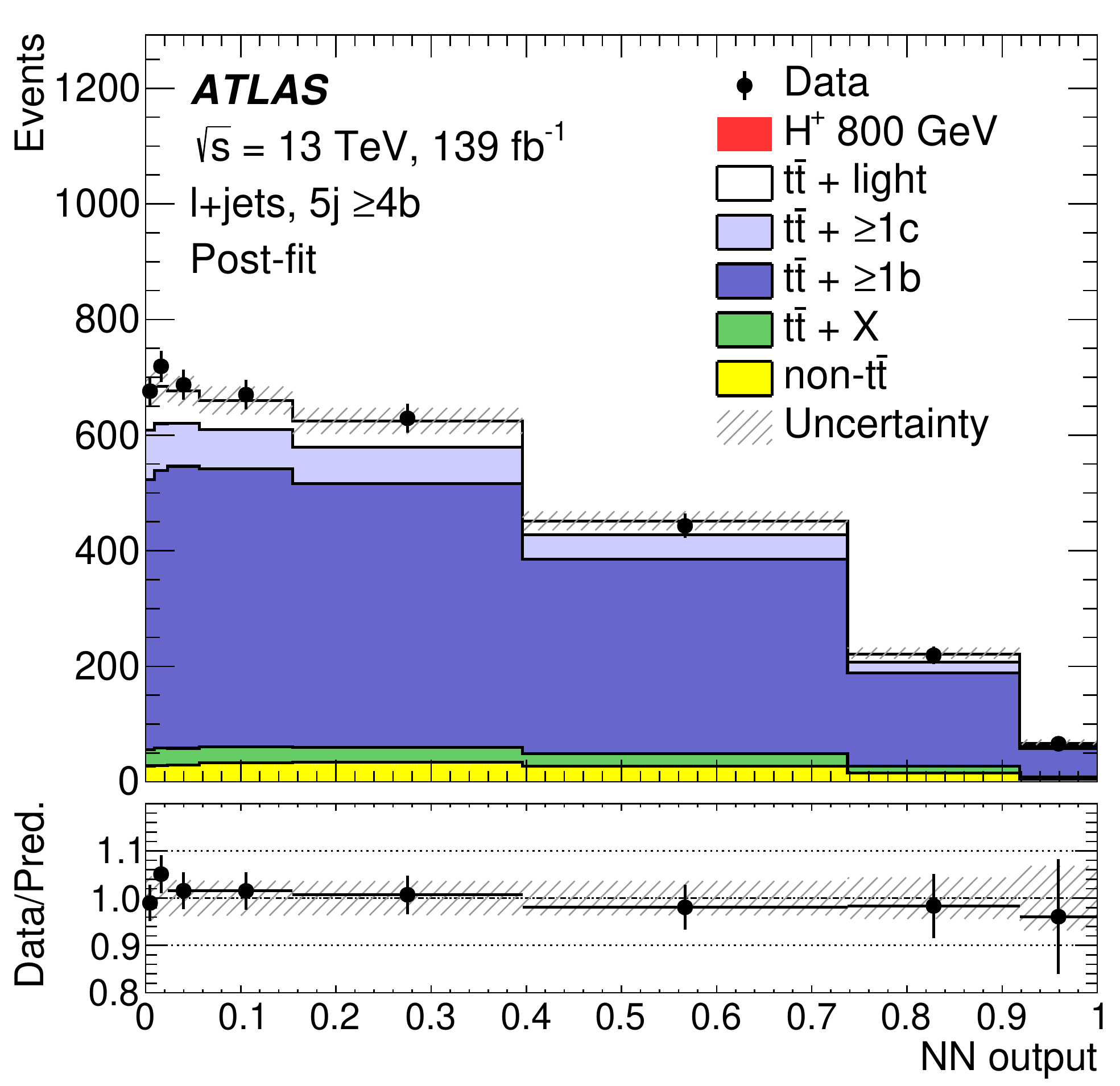}}
\subfloat[$\geq$ 6j3b]
{\includegraphics[width = 0.25\textwidth]{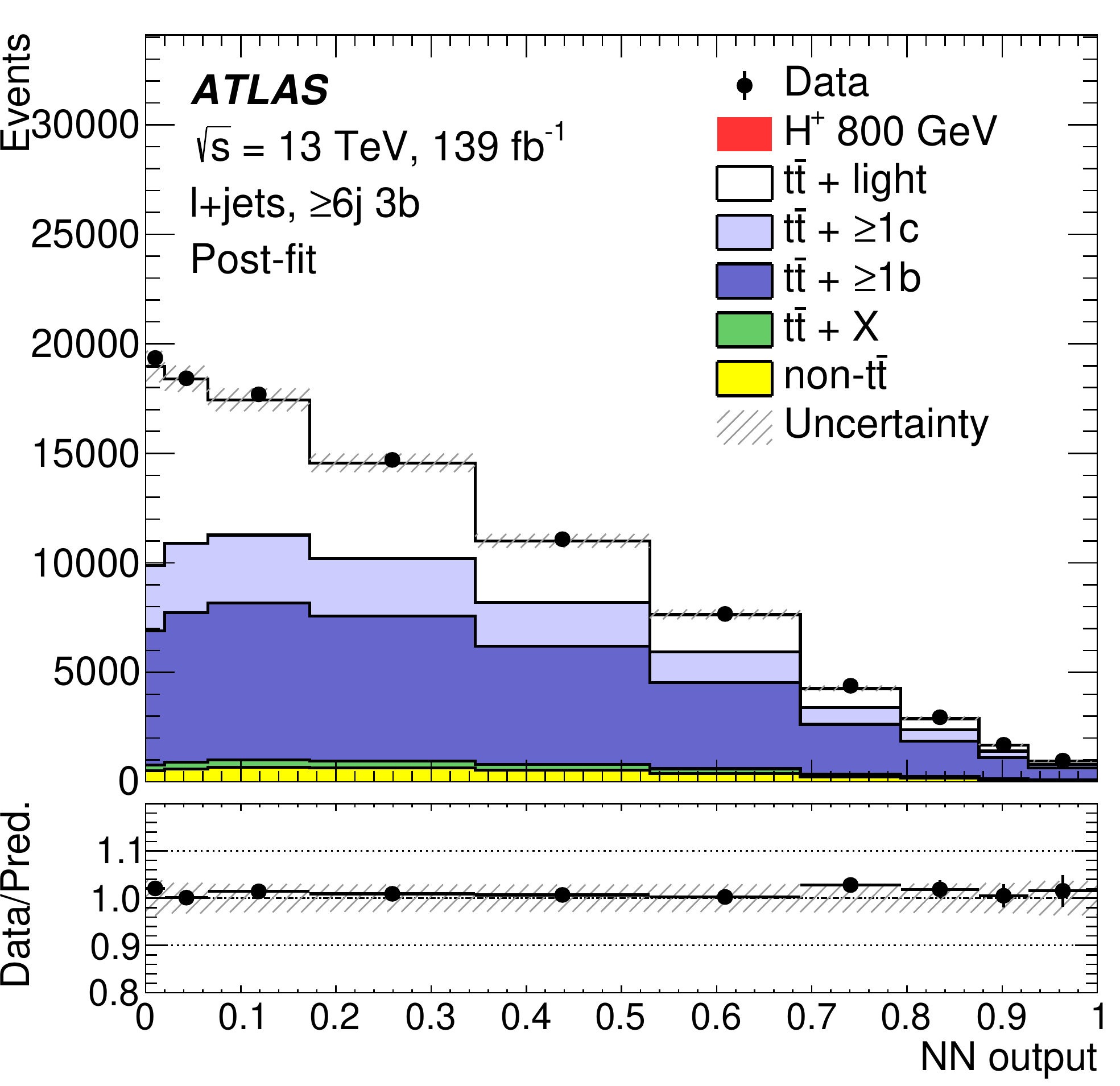}}
\subfloat[$\geq$6j$\geq$4b]
{\includegraphics[width = 0.25\textwidth]{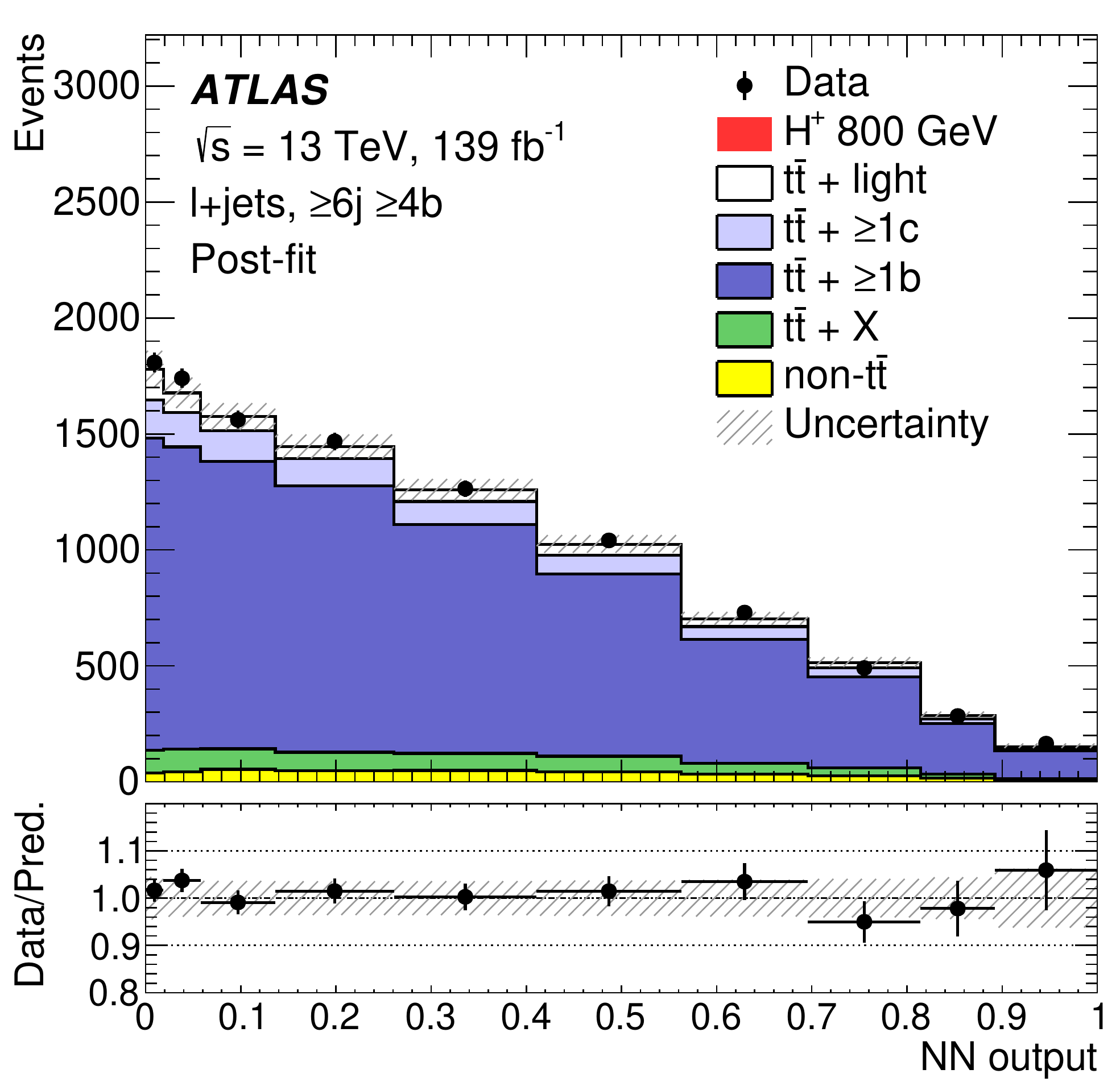}} \\
\caption{Distributions of the NN output after the fit for the 200~\GeV\ (top) and
800~\GeV\ (bottom) \hplus\ mass hypotheses in the four analysis regions.
The lower panels display the ratio of the data to the total prediction.
The hatched bands show the uncertainties after the fit.
}
\label{fig:NNpostfit}
\end{figure}

The 95\% confidence level (CL) upper limits on $\sigma(pp\rightarrow\tbhplus) \times \BR(\hplustb)$
obtained using the CL$_{\textrm S}$ method are presented in Figure~\ref{fig:upperlimit}.
Uncertainties in the predicted \hplus\ cross-sections or branching ratios are not included.
The observed (expected) limits range from
$\sigma \times\BR = 3.6\,(2.6)$~pb at $\mhp = 200$~\GeV\ to
$\sigma \times\BR = 0.036\,(0.019)$~pb at $\mhp = 2$~\TeV.
 
\begin{figure}[htb]
\centering
{\includegraphics[width = 0.9\textwidth]{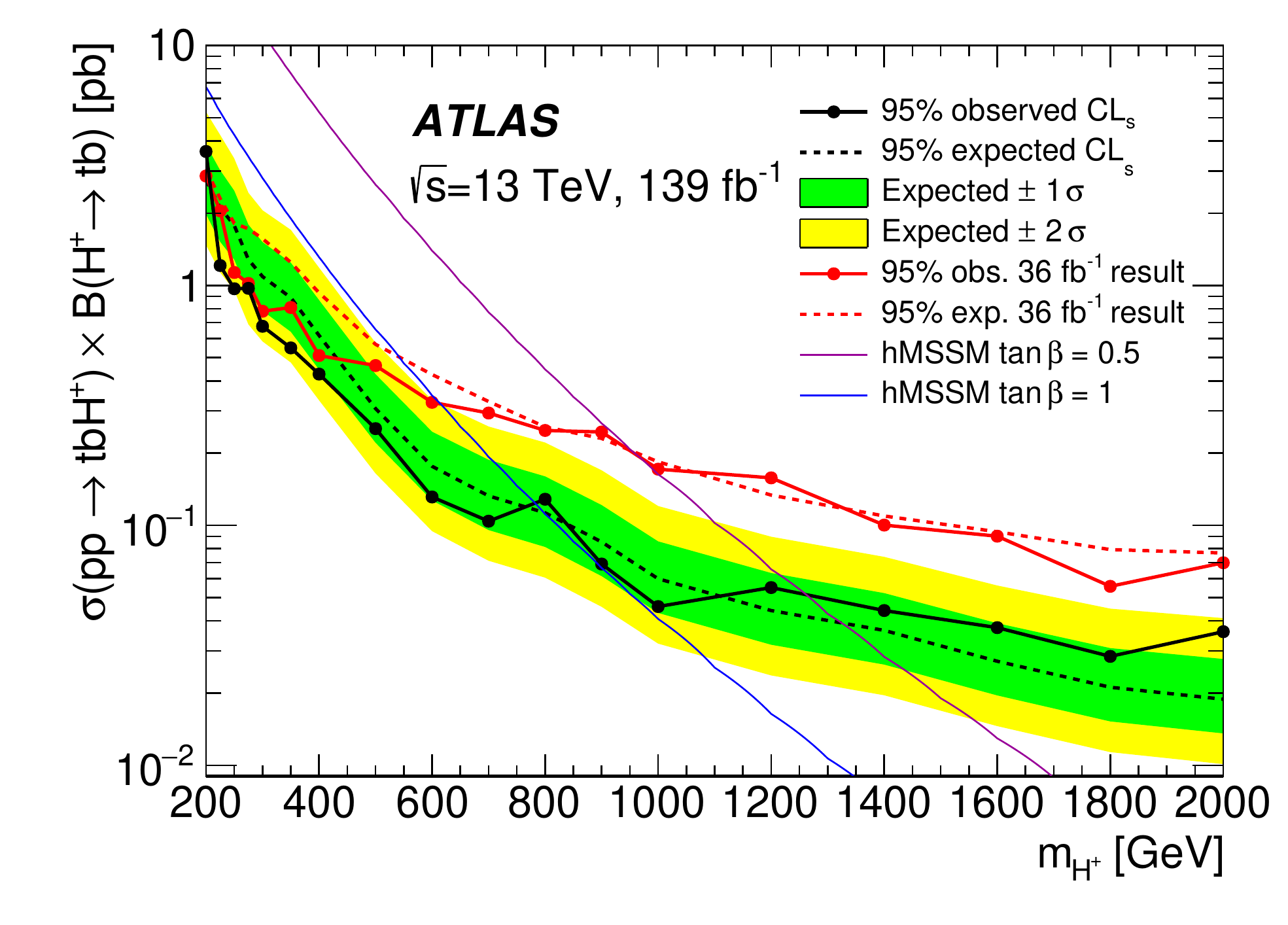}}
\caption{Observed and expected upper limits for the production of \hplustb\ in association
with a top quark and a bottom quark.
The bands surrounding the expected limit show the 68\% and 95\% confidence intervals.
The red lines show the observed and expected 95\% CL exclusion limits obtained with
the 36~\ifb\ data sample~\cite{HIGG-2017-04}. 
Theory predictions are shown for two representative values of $\tan\beta$ in the hMSSM benchmark scenario.
Uncertainties in the predicted \hplus\ cross-sections or branching ratios are not considered.}
\label{fig:upperlimit}
\end{figure}
 
Figure~\ref{fig:exclusion} shows 95\% CL exclusion limits set on $\tan\beta$
as a function of \mhp\ for various benchmark scenarios in the MSSM. It is the first time that they are shown for all $M_h^{125}$ available scenarios using the $\hplustb$ channel. In the hMSSM framework,
effective couplings of the lighter Higgs boson to the top quark,
bottom quark and vector bosons are derived from fits to LHC data on the production and decay rates of the observed Higgs boson,
including the limits from the search for heavier neutral and charged Higgs boson states.
The $M_h^{125}$, $M_h^{125}(\tilde{\chi})$, $M_h^{125}(\tilde{\tau})$, $M_h^{125}(\mathrm{alignment})$ and
$M_{h_1}^{125}(\mathrm{CPV})$
scenarios also feature a scalar particle with mass and couplings compatible with those of the observed Higgs boson,
and force a significant portion of their parameter space to be compatible with the limits from searches
for supersymmetric particles.
In the $M_h^{125}$ scenario, all supersymmetric particles are relatively heavy and the decays of the MSSM Higgs bosons are essentially unaffected,
whereas the $M_h^{125}(\tilde{\chi})$ and $M_h^{125}(\tilde{\tau})$ models include either light charginos and neutralinos ($M_h^{125}(\tilde{\chi})$) or light staus ($M_h^{125}(\tilde{\tau})$).
In both cases a charged Higgs boson of sufficiently high mass is allowed to decay into the supersymmetric particles.
Finally, the value of $\tan\beta$
in both the $M_h^{125}(\mathrm{alignment})$ scenario,
characterised by one of the two neutral CP-even scalars having couplings like those of the SM Higgs boson,
and the $M_{h_1}^{125}(\mathrm{CPV})$ scenario,
which includes CP violation in the Higgs sector,
is already constrained to be in the range 1--20 by previous searches at the LHC~\cite{Bagnaschi_2019}.
Uncertainties in the predicted \hplus\ cross-sections or branching ratios are not included in the limits.
For all scenarios except the hMSSM,
Higgs boson masses and mixing (and effective Yukawa couplings)
have been calculated with the code
FeynHiggs~\cite{Heinemeyer:1998yj, Heinemeyer:1998np, Degrassi:2002fi, Frank:2006yh, Hahn:2013ria, Bahl:2016brp, Bahl:2017aev}.
Whereas in the hMSSM the branching ratios are computed solely with
HDECAY~\cite{Djouadi:1997yw, Djouadi:2018xqq},
all other scenarios combine the most precise results of
FeynHiggs, HDECAY and PROPHECY4f~\cite{Bredenstein:2006rh, Bredenstein:2006ha}.
 
In the context of these scenarios,
$\tan\beta$ values below 1 are observed to be excluded at 95\% CL for \hplus\ masses between 200 and $\sim$790~\GeV.
High values of $\tan\beta$ between 34 and 60 are excluded in a similar mass range in the hMSSM and $M_h^{125}(\tilde{\chi})$ models.
The most stringent limit, $\tan\beta < 2.1$ excluded at 95\% CL,
is set for the \hplus\ mass hypothesis of 225~\GeV\ in the hMSSM and for the 250~\GeV\ \hplus\ mass hypothesis in the
$M_h^{125}$, $M_h^{125}(\tilde{\chi})$, $M_h^{125}(\tilde{\tau})$, $M_h^{125}$(alignment) and $M_{h_1}^{125}$(CPV) models.
The low $\tan\beta$ and high \hplus\ mass parameter space was not excluded by any other analysis before, while the high
$\tan\beta$ was already excluded by the $\hplus\rightarrow\tau\nu$ search. Compared to previous results of the same search channel,
this analysis excludes a broader region of large $\tan\beta$. Additionally, an extended region of low $\tan\beta$ and low and high \hplus\ masses is also excluded.

\begin{figure}[htb]
\centering
\subfloat[]
{\includegraphics[width = 0.36\textwidth]{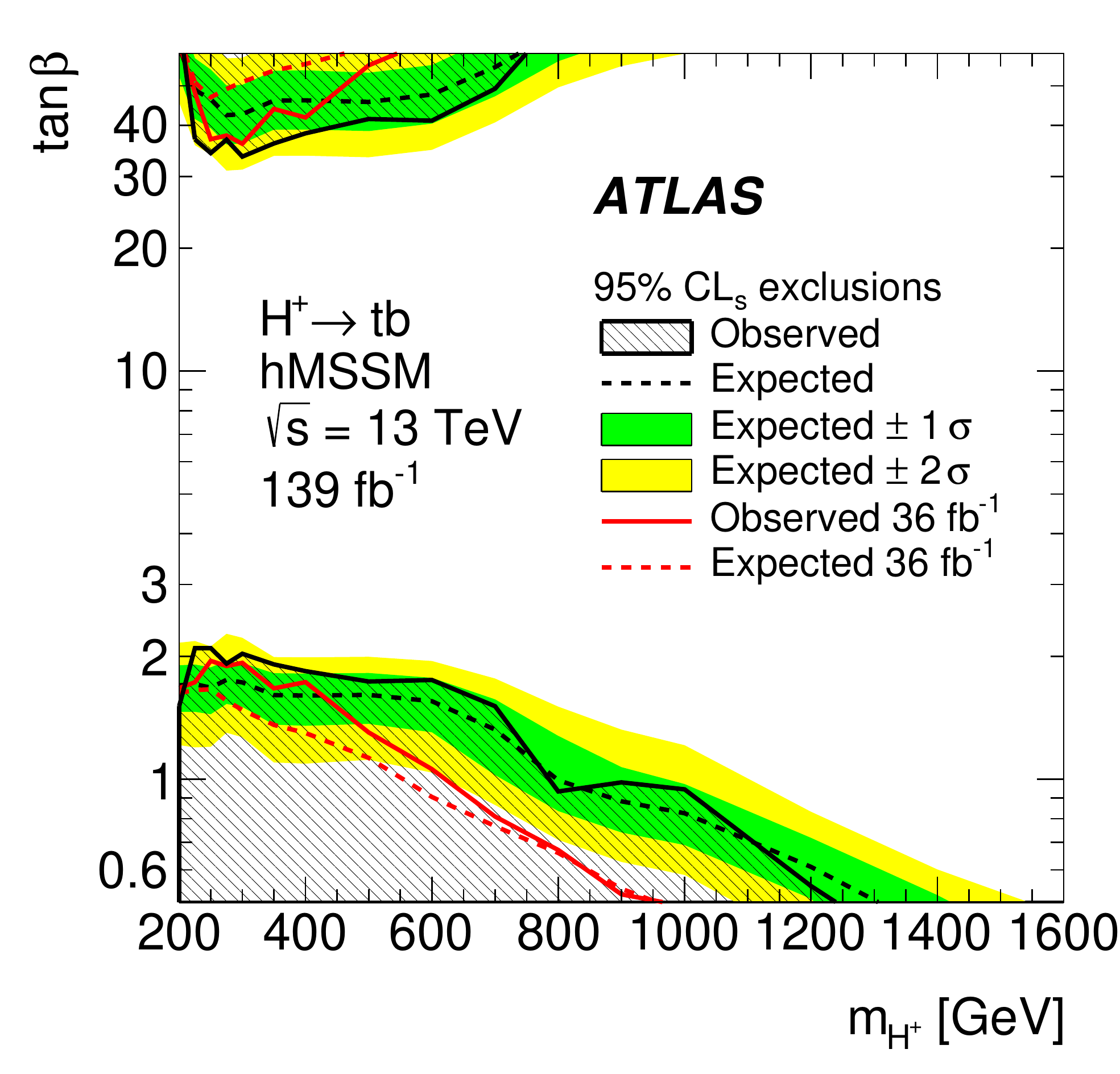}}
\subfloat[]
{\includegraphics[width = 0.36\textwidth]{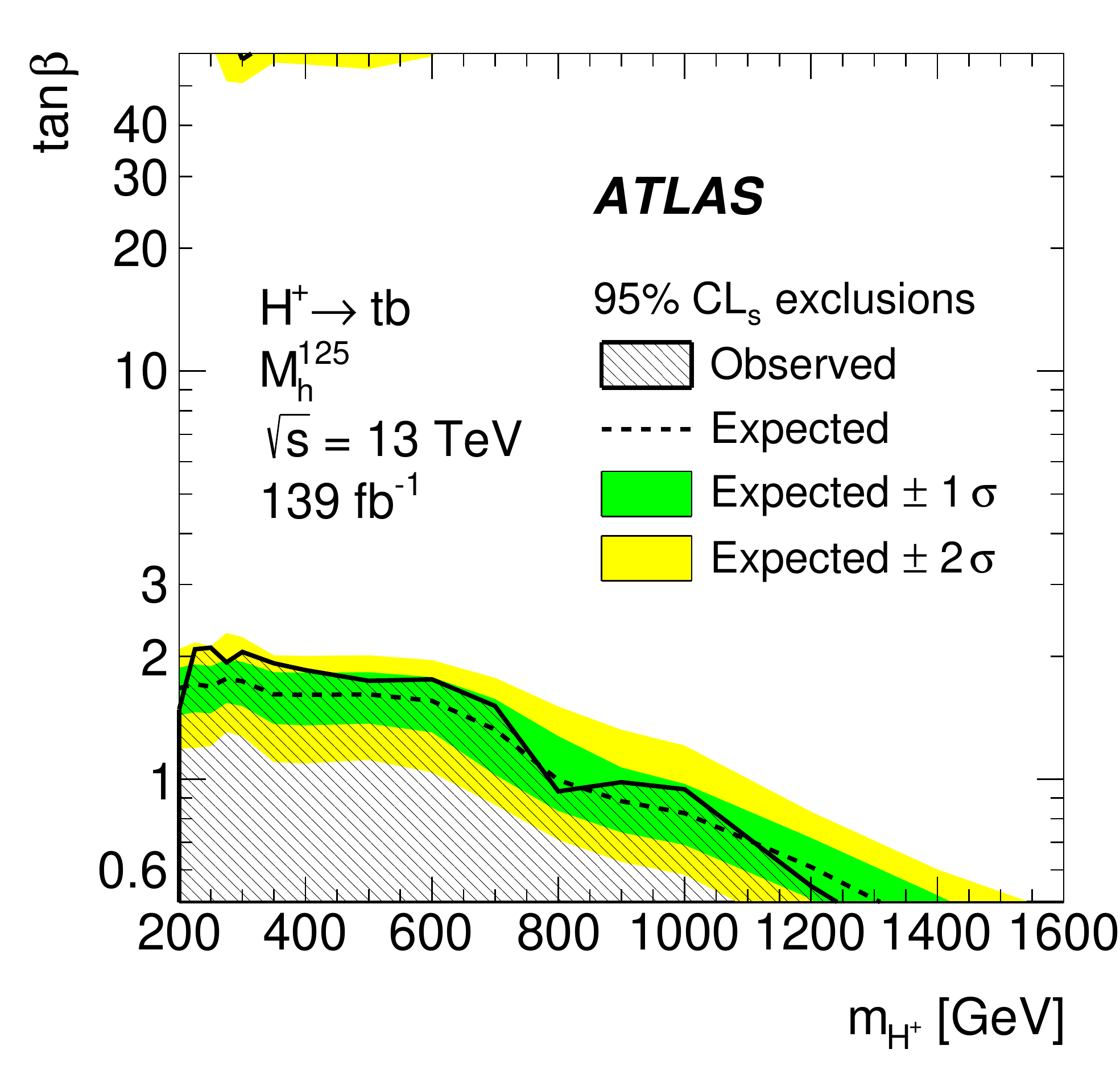}} \\
\subfloat[]
{\includegraphics[width = 0.36\textwidth]{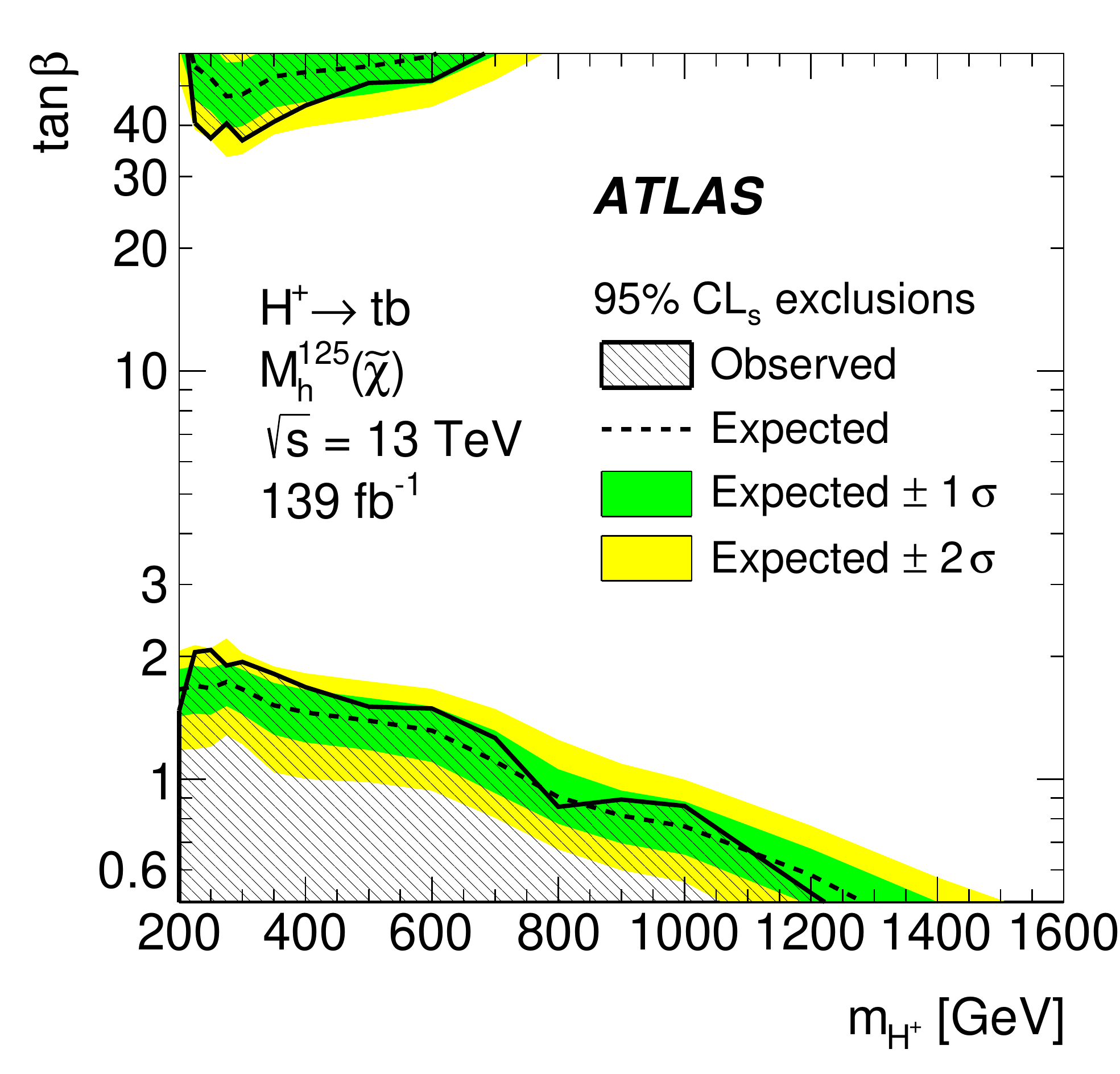}}
\subfloat[]
{\includegraphics[width = 0.36\textwidth]{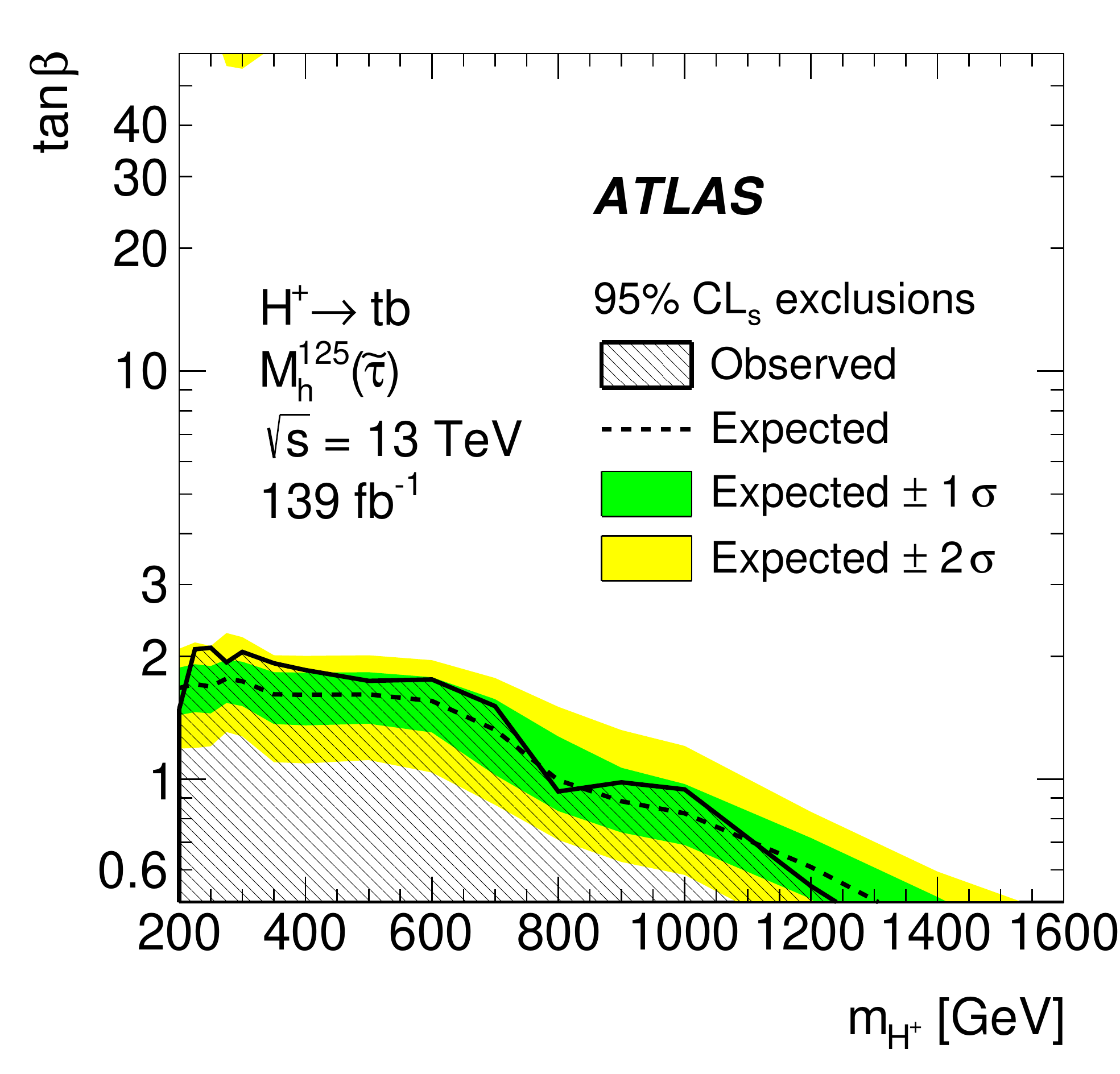}} \\
\subfloat[]
{\includegraphics[width = 0.36\textwidth]{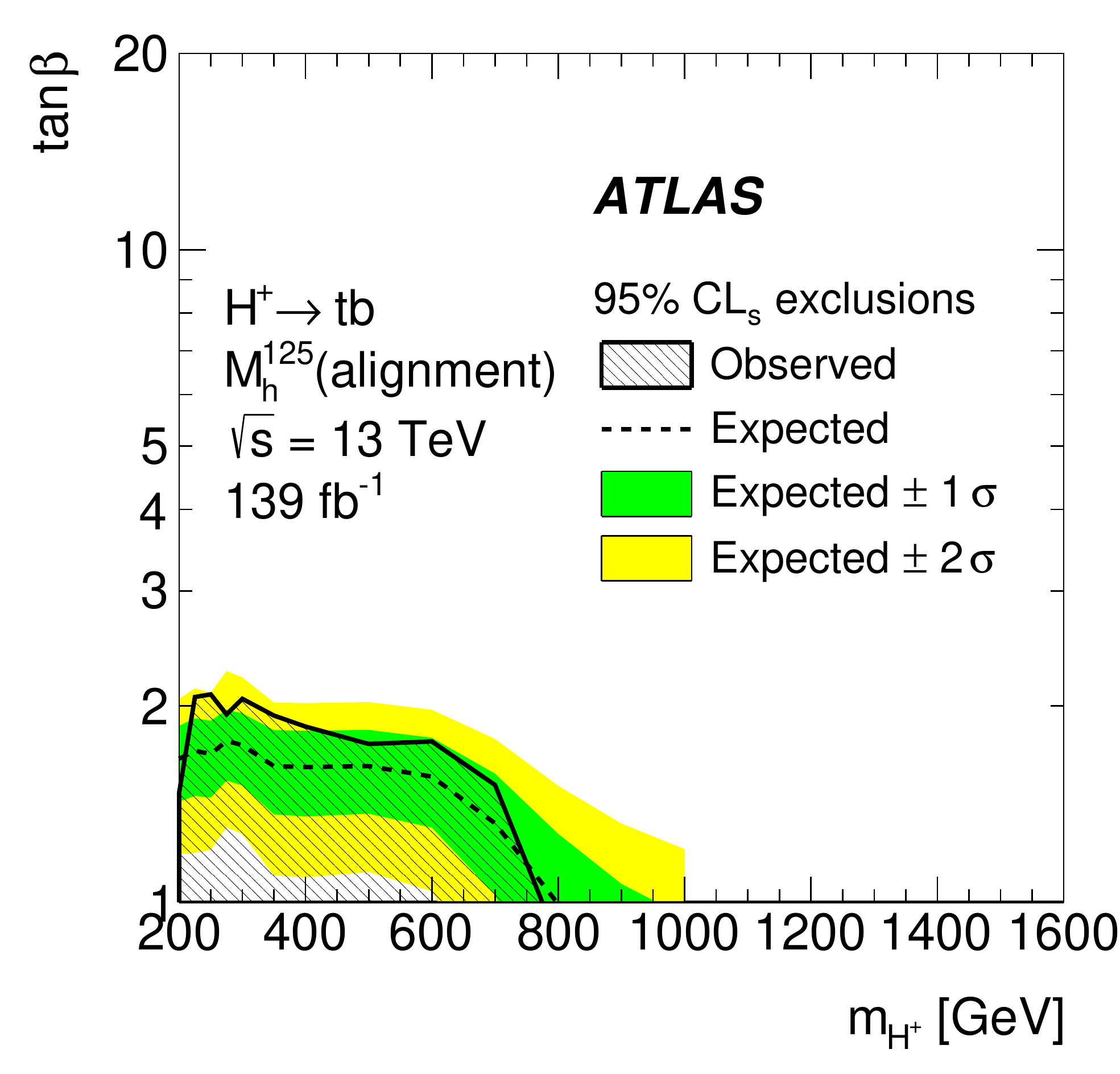}}
\subfloat[]
{\includegraphics[width = 0.36\textwidth]{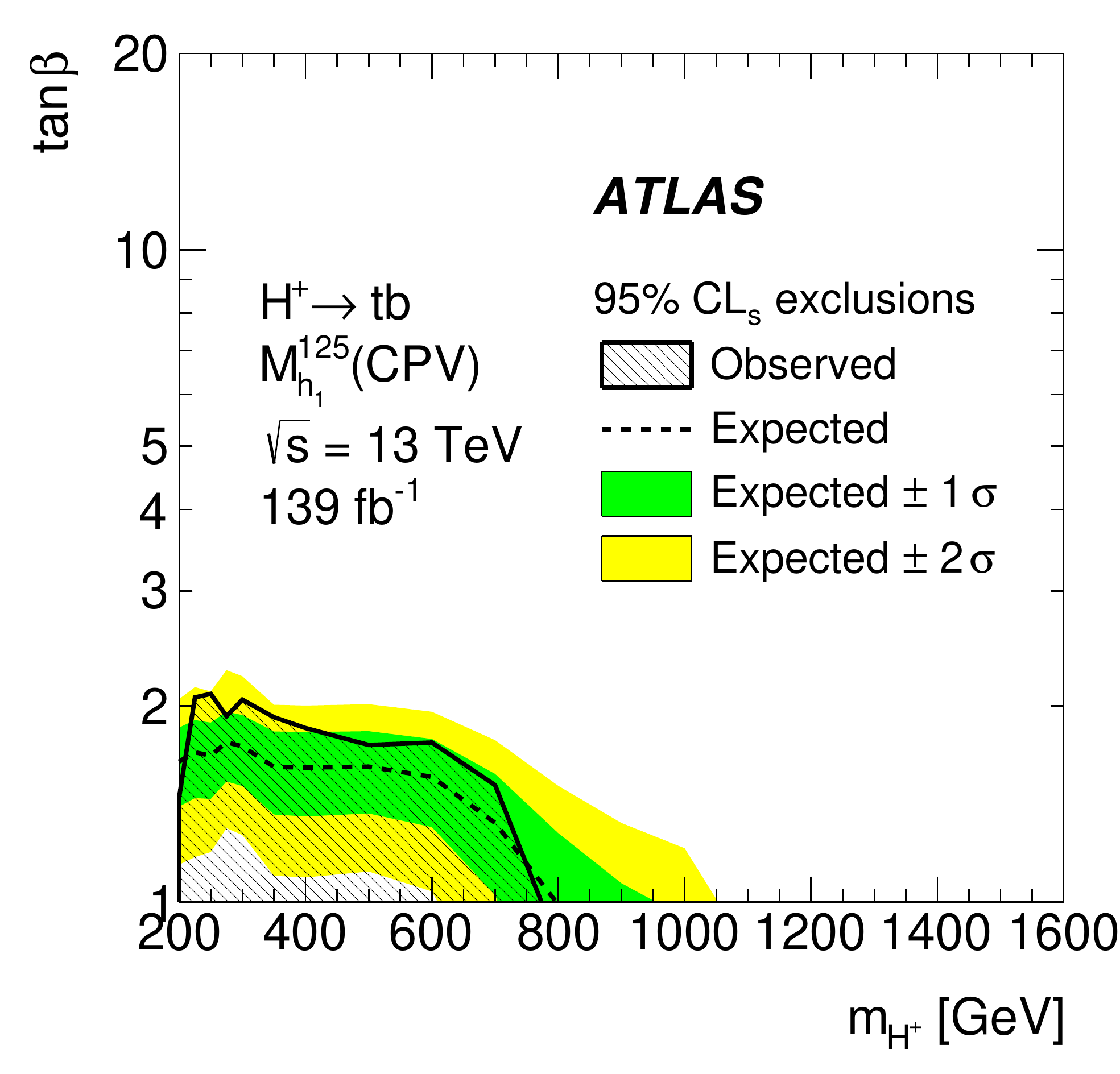}} \\
\caption{Observed and expected limits on $\tan\beta$ as a function of \mhp\ in various scenarios:
(a) hMSSM, (b) $M_h^{125}$, (c) $M_h^{125}(\tilde \chi)$, (d) $M_h^{125}(\tilde\tau)$, (e) $M_h^{125}$(alignment) and (f) $M_{h_1}^{125}$(CPV).
Limits are shown for $\tan\beta$ values in the range of 0.5--60 or 1--20 depending on the availability of model predictions.
The bands surrounding the expected limits show the 68\% and 95\% confidence intervals.
Uncertainties in the predicted \hplus\ cross-sections or branching ratios are not considered.}
\label{fig:exclusion}
\end{figure}

\FloatBarrier

\section{Conclusion}
A search for charged Higgs bosons is presented using a data sample corresponding to
an integrated luminosity of 139~\ifb\ from proton--proton collisions at \comEne,
recorded with the ATLAS detector at the LHC.
The search for $pp\rightarrow\tbhplus$ is performed in the \hplus\ mass
range 200--2000~\GeV.
A neural network that combines jet multiplicities and several kinematic variables is built in the regions where the signal rate is expected to be largest.
The neural network output depends on the \hplus\ mass,
and a fit to the data is performed simultaneously on the neural network output distributions in the analysis regions,
separately for each signal mass hypothesis.
 
No significant excess above the expected Standard Model background is found and observed (expected)
upper limits at 95\% confidence level are set on the $\sigma (pp\rightarrow \tbhplus)$ production cross-section times the
branching ratio $\BR(\hplustb)$,
which range from $\sigma \times \BR  = 3.6\, (2.6)$~pb at $\mhp = 200$~\GeV\
to $\sigma \times \BR  = 0.036\,(0.019)$~pb at $\mhp = 2$~\TeV.
Compared to the previous ATLAS search for \tbhplus\ production followed by \hplustb\ decays with 36~\ifb,
the observed $\sigma \times \BR$ limits improved by 5\% to 70\%,
depending on the \hplus\ mass, apart from the lowest one.
In the low \hplus\ mass region the measurement is dominated by systematic uncertainties,
while in the high \hplus\ mass region the use of tighter lepton triggers and refined $b$-tagging techniques,
along with the \hplus\ mass-independent training of the neural network,
leads to an improvement beyond the simple scaling with the square root of the ratio of integrated luminosities.
 
In the context of the hMSSM and several  $M_h^{125}$ scenarios,
some values of $\tan\beta$,
in the range 0.5--2.1,
are excluded for \hplus\ masses between 200 and 1200~\GeV.
For \hplus\ masses between $\sim$200 and $\sim$750~\GeV,
values of $\tan\beta >$ 34 are also excluded.
Compared to previous results of the same search channel,
this analysis excludes a broader region of large $\tan\beta$.
Additionally, an extended region of low $\tan\beta$ and low and high \hplus\ masses is also excluded.
 
\section*{Acknowledgements}  
 

We thank CERN for the very successful operation of the LHC, as well as the
support staff from our institutions without whom ATLAS could not be
operated efficiently.
 
We acknowledge the support of ANPCyT, Argentina; YerPhI, Armenia; ARC, Australia; BMWFW and FWF, Austria; ANAS, Azerbaijan; SSTC, Belarus; CNPq and FAPESP, Brazil; NSERC, NRC and CFI, Canada; CERN; ANID, Chile; CAS, MOST and NSFC, China; COLCIENCIAS, Colombia; MSMT CR, MPO CR and VSC CR, Czech Republic; DNRF and DNSRC, Denmark; IN2P3-CNRS and CEA-DRF/IRFU, France; SRNSFG, Georgia; BMBF, HGF and MPG, Germany; GSRT, Greece; RGC and Hong Kong SAR, China; ISF and Benoziyo Center, Israel; INFN, Italy; MEXT and JSPS, Japan; CNRST, Morocco; NWO, Netherlands; RCN, Norway; MNiSW and NCN, Poland; FCT, Portugal; MNE/IFA, Romania; JINR; MES of Russia and NRC KI, Russian Federation; MESTD, Serbia; MSSR, Slovakia; ARRS and MIZ\v{S}, Slovenia; DST/NRF, South Africa; MICINN, Spain; SRC and Wallenberg Foundation, Sweden; SERI, SNSF and Cantons of Bern and Geneva, Switzerland; MOST, Taiwan; TAEK, Turkey; STFC, United Kingdom; DOE and NSF, United States of America. In addition, individual groups and members have received support from BCKDF, CANARIE, Compute Canada, CRC and IVADO, Canada; Beijing Municipal Science \& Technology Commission, China; COST, ERC, ERDF, Horizon 2020 and Marie Sk{\l}odowska-Curie Actions, European Union; Investissements d'Avenir Labex, Investissements d'Avenir Idex and ANR, France; DFG and AvH Foundation, Germany; Herakleitos, Thales and Aristeia programmes co-financed by EU-ESF and the Greek NSRF, Greece; BSF-NSF and GIF, Israel; La Caixa Banking Foundation, CERCA Programme Generalitat de Catalunya and PROMETEO and GenT Programmes Generalitat Valenciana, Spain; G\"{o}ran Gustafssons Stiftelse, Sweden; The Royal Society and Leverhulme Trust, United Kingdom.
 
The crucial computing support from all WLCG partners is acknowledged gratefully, in particular from CERN, the ATLAS Tier-1 facilities at TRIUMF (Canada), NDGF (Denmark, Norway, Sweden), CC-IN2P3 (France), KIT/GridKA (Germany), INFN-CNAF (Italy), NL-T1 (Netherlands), PIC (Spain), ASGC (Taiwan), RAL (UK) and BNL (USA), the Tier-2 facilities worldwide and large non-WLCG resource providers. Major contributors of computing resources are listed in Ref.~\cite{ATL-SOFT-PUB-2020-001}.
 

\clearpage

\printbibliography

\clearpage 
 
\begin{flushleft}
\hypersetup{urlcolor=black}
{\Large The ATLAS Collaboration}

\bigskip

\AtlasOrcid[0000-0002-6665-4934]{G.~Aad}$^\textrm{\scriptsize 102}$,    
\AtlasOrcid[0000-0002-5888-2734]{B.~Abbott}$^\textrm{\scriptsize 128}$,    
\AtlasOrcid[0000-0002-7248-3203]{D.C.~Abbott}$^\textrm{\scriptsize 103}$,    
\AtlasOrcid[0000-0002-2788-3822]{A.~Abed~Abud}$^\textrm{\scriptsize 36}$,    
\AtlasOrcid[0000-0002-1002-1652]{K.~Abeling}$^\textrm{\scriptsize 53}$,    
\AtlasOrcid[0000-0002-2987-4006]{D.K.~Abhayasinghe}$^\textrm{\scriptsize 94}$,    
\AtlasOrcid[0000-0002-8496-9294]{S.H.~Abidi}$^\textrm{\scriptsize 167}$,    
\AtlasOrcid[0000-0002-8279-9324]{O.S.~AbouZeid}$^\textrm{\scriptsize 40}$,    
\AtlasOrcid{N.L.~Abraham}$^\textrm{\scriptsize 156}$,    
\AtlasOrcid[0000-0001-5329-6640]{H.~Abramowicz}$^\textrm{\scriptsize 161}$
% ...
\bigskip
\\

% ...
$^{36}$CERN, Geneva; Switzerland.\\
% ...
$^{40}$Niels Bohr Institute, University of Copenhagen, Copenhagen; Denmark.\\
% ...
$^{53}$II. Physikalisches Institut, Georg-August-Universit\"{a}t G\"ottingen, G\"ottingen; Germany.\\
% ...
$^{94}$Department of Physics, Royal Holloway University of London, Egham; United Kingdom.\\
% ...
$^{102}$CPPM, Aix-Marseille Universit\'e, CNRS/IN2P3, Marseille; France.\\
$^{103}$Department of Physics, University of Massachusetts, Amherst MA; United States of America.\\
% ...
$^{128}$Homer L. Dodge Department of Physics and Astronomy, University of Oklahoma, Norman OK; United States of America.\\
% ...
$^{156}$Department of Physics and Astronomy, University of Sussex, Brighton; United Kingdom.\\
% ...
$^{161}$Raymond and Beverly Sackler School of Physics and Astronomy, Tel Aviv University, Tel Aviv; Israel.\\
% ...
$^{167}$Department of Physics, University of Toronto, Toronto ON; Canada.\\
% ...

\end{flushleft}


\end{document}